# Experimental and theoretical Brownian Dynamics analysis of ion transport during cellular electroporation of *E. coli* bacteria

Juan A. González-Cuevas[1*], Ricardo Argüello[1], Marcos Florentin[2], Franck M. André[3] and Lluis M. Mir[3]

[1]School of Engineering, National University of Asunción, Campus San Lorenzo 2169, Paraguay
[2]School of Chemistry, National University of Asunción, Campus San Lorenzo 2169, Paraguay
[3]Université Paris-Saclay, CNRS, Gustave Roussy, UMR 9018 METSY, 94805 Villejuif, France
*Corresponding author: jgonzalez@ing.una.py

## Abstract

*Escherichia coli* bacteria is a rod-shaped organism composed of a complex double membrane structure. Knowledge of electric field driven ion transport through both membranes and the evolution of their induced permeabilization has important applications in biomedical engineering, delivery of genes and antibacterial agents. However, few studies have been conducted on Gram negative bacteria in this regard considering the contribution of all ion types. To address this gap in knowledge, we have developed a deterministic and stochastic Brownian dynamics model to simulate in 3-D space the motion of ions through pores formed in the plasma membranes of *Escherichia coli* cells during electroporation. The diffusion coefficient, mobility and translation time of $Ca^{2+}$, $Mg^{2+}$, $Na^+$, $K^+$ and $Cl^-$ ions within the pore region are estimated from the numerical model. Calculations of pore's conductance have been validated with experiments conducted at Gustave Roussy. From the simulations, it was found that the main driving force of ionic uptake during the pulse is the one due to the externally applied electric field. The results from this work provide a better understanding of ion transport during electroporation, aiding in the design of electrical pulses for maximizing ion throughput, primarily for application in cancer treatment.

**Keywords:** electroporation, ion transport, pore conductivity, cancer treatment, membrane crossing time, diffusion coefficient, mobility, *E. coli* bacteria

# INTRODUCTION

Electroporation consists in the application of an electric field to biological cells that creates aqueous pathways or pores in the membrane with an estimated minimum radius of 1 nm. The applied electric field causes an induced transmembrane voltage superimposed on the resting membrane potential present under physiological conditions, reaching voltages in the 0.2 –1 V range [68] [26]. Electroporation is a subset of electropermeabilization, which ascribes the electrically induced increase in the membrane permeability to a broader range of physical and chemical changes to the membrane lipid bilayer and modulation of its proteins' function [30].

Most biomedical applications of electroporation aim to transport extracellular ions or molecules into cells, such as calcium electroporation, wherein supraphysiological doses of calcium are internalized causing cell death [27], irreversible electroporation (IRE) [44], which is a non-thermal tissue ablation therapy, electrochemotherapy (ECT), consisting of established anti-cancer drugs with added charge groups to prevent spontaneous entry [21], and non-viral electro-gene transfer, where DNA is first introduced into a tissue, possibly by needle injection, and then the tissue's cells are electroporated within the desired treatment volume [54] [62]. Thus, improving the fundamental understanding of ionic transport across the permeabilized membrane, such as diffusion, mobility, and translation times remains a priority given the importance of these phenomena in cell biology and medical applications.

Pulse duration for conventional electroporation is on the order of or longer than the membrane charging time, which also has some impact on the transmembrane potential. Higher-field (> 1 kV/cm), longer pulses ( > > 1 ms) are involved in IRE, necrosis, and electrocution injury [69]. Lower-field (< 1kV/cm), shorter pulses (between 100 µs and 1 ms) are used in ECT and gene delivery [69]. High-field short duration pulses have a negligible thermal effect on the cells and tissues and can preferentially target intracellular components. However, if the field is too high in nanosecond pulses ( > 20 kV/cm) apoptosis occurs [69]. Consequently, it is possible to tailor the operating pulse width to achieve preferential energy deposition into organelles based on their geometry, size, and location.

Manipulation of cells via electroporation can be done *in vivo, ex vivo and in vitro*. The present work is focused on modeling ion transport during electroporation using *in vitro* data of Gram-negative *Escherichia coli* (*E. coli*) bacteria for validation. The DH5a strain of *E. coli* bacteria is used in our research because it is a non-pathogenic facultative anaerobe which is easy and inexpensive to grow in large quantities in simple culture media. It is a well-characterized bacteria that is easy to obtain and isolate from humans and animals. It provides insights into the physiology of other pathogenic bacteria of the same family, which can aid in the development of treatments and pharmaceutical products [30]. Due to the difference in anatomy between bacterial and mammalian cells, in particular the double membrane structure with a periplasm region in *E coli*, vastly higher electric fields need to be applied for electroporation, on the order of tens of kV/cm. While the membrane structure is more complex in *E. coli* than in mammalian cells, *E. coli* is a simpler organism overall since it is a prokaryote which lacks a nucleus and other organelles [30].

Molecular Dynamics (MD) is a microscopic simulation scheme which is highly representative of the molecular-level structure of the membrane. It Is well suited to simulate pore formation considering the details of the chemical components of a small region of the lipid bilayer, as performed by Piggot et. al. [50] for S. Aureus and the outer membrane of *E. coli.* However, long time scales are required to corroborate experimental conductivity measurements, which cannot be obtained with microscopic MD simulations due to their long computation time [11]. Besides, the atomistic structure of pores is not fully known experimentally, and inaccurate models of water molecules may be used in the calculations [51] [39]. On the other hand, macroscopic analyses that treat the cellular matter as a continuum fluid with voltage barriers are more suitable for longer time scales but are less representative of the biological structure of the membrane's pore [56] [55].

Therefore, simulations of ionic transport of electroporated *E. coli* bacteria were carried out in the present work using a semi-microscopic Brownian Dynamics (BD) scheme, which combines the advantages of both microscopic and macroscopic models, but also inherits some of their disadvantages, namely time step dependence, exclusion of chemical reactions, loss of information due to simplified dynamics, and accumulation of errors over time [5]. Besides, BD models are Markov processes that are limited to steady state dilute systems [9]. However, the advantages outweigh the disadvantages considering the characteristics of the system being studied. For instance, the BD approach is computationally more efficient and faster than MD simulations, since it is a coarse-grained model that groups collectively multiple membrane and water molecules [51] [39]. This allows for the study at longer time scales of a larger system consisting of both the inner and outer membranes of *E. coli*, while keeping a higher resolution than macroscopic continuum models [51] [39]. The BD scheme incorporates hydrodynamic interactions and thermal fluctuations directly through stochastic forces, allowing for longer time steps ΔT and requiring fewer samples to capture the essential statistical behavior of the system [51] [39]. The intermediate resolution level of BD simulations facilitates multiscale bridging between atomistic and macroscopic phenomena [39], as has been done in the present work for conductivity validation. Thus, MD simulations are suitable to study biomolecular dynamics in solutions, diffusion transport processes and ionic conductance of channels and pores [51].

## MATERIALS AND METHODS

### *Experimental Setup*

Experimental studies were conducted at the Vectorology Department of the Gustave Roussy Institute using DH5a *E. Coli* bacteria in order to obtain measurements of conductivity and permeability of cell membranes after electroporation. In this regard, a Bio-Rad Gene Pulser Xcell™ electroporator set to voltage $V_0$ = 1800 V, resistance $R$ = 150 Ω and capacitance $C$ = 50 μF was used to apply a single exponential decay electrical pulse with time constant $\tau = RC = 7.5\ ms$ on a 1mL capacity Bio-Rad plastic cuvette with gap width $w$ = 1 mm, creating over time t an electric field $\vec{E} = V_0 e^{-t/RC}/w$ = 18 kV/cm at $t$ = 0.

Figure 1 a) shows the experimental arrangement with a simplified circuitry of the pulse generator. The Bio-Rad cuvette incorporates aluminum electrodes on two opposite internal walls which get in contact with the electrical circuit. Switch 1 is first closed to charge the capacitor and then opened to isolate it from the source. The pulse is discharged on the cuvette suspension by closing switch 2. Figure 1 b) represents the waveform of the applied pulse. The amount of current released by the capacitor is related to the time at which the energy stored in it is allowed to dissipate, which depends on the resistance across the electrode assembly and sample. Therefore, the higher the resistance, the longer the pulse. Since the *E. coli* cell charging time is usually 100 ns, long pulses must be used if the membrane potential is to exceed 0.5-1V for permeabilization.

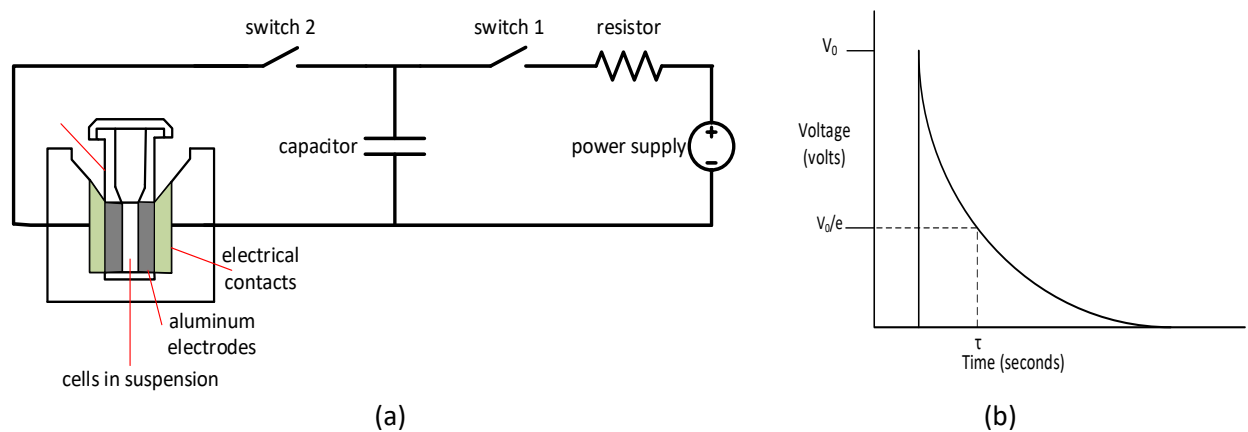

**FIGURE 1**. a) Simplified electrical circuit diagram of the electroporator device. b) Applied unipolar exponential decay waveform with time constant τ = RxC

Conductivity assays were conducted in three different physiological states: fresh bacteria in growth phase, overnight bacteria coming from a saturated culture, and electrocompetent bacteria. *E. coli* DH5a was grown in Luria broth (LB) media. Cultures were kept at 37 ºC, shaking at 200 rpm. For the electroporation selection, LB agar plates were prepared using 50 µg/mL kanamycin. Electrocompetent *E. coli* were prepared by growing the bacteria in 500 mL of LB media at 200 rpm until reaching an optical density of 0.4 – 0.6. The whole volume was then centrifuged at 3000 rpm at 4 ºC for 15 min, discarding the supernatant and resuspending the pellet in 500 mL 10% glycerol in DI water. This process was repeated thrice by resuspending the cells into 200 mL, 20 mL and 2 mL, aliquoting the final volume into 100 uL fractions.

Before conducting each conductivity assay, the FiveEasy FiveGo FE30 conductometer probe was washed with deionized (DI) water, then 70% ethanol in DI water, then washed thrice with DI water. This step was necessary to reduce conductivity to background levels (1.5 µS/cm or less). A special column for measuring a minimum amount of sample was crafted by cutting a disposable 25 mL plastic pipette of enough diameter to fit the conductometer's probe, which was then attached to a flat acrylic base. This column was also washed as described. In order to perform triplicate assays, 33 uL of each sample, or 33 uL of sample plus 1 uL of 1 µM YO-PRO-1 were taken into a final volume of 1500 DI water.

Flow cytometry assays were performed on a BD Accuri™ C6 Flow Cytometer in order to determine the cell count and differences in cell membrane permeabilization in the cited states. A 488 nm laser was used for YO-PRO-1 excitation. The capillary was cleaned by flowing DI water before starting. The device was blanked with 233 µL DI water, 10% glycerol in DI water or 10% LB in DI water, depending on the sample. 233 µL of sample dilution were then analyzed on the Fast delivery rate at 66 µL/min, by diluting 33 uL of each sample with 200 uL DI water or the corresponding diluent. For the samples tested with the nucleic acid stain, 1 µL of 1 mM YO-PRO-1 was added to 33 µL of sample plus 199 µL of DI water or the corresponding diluent.

Prior to the electroporation experiment the bacterial samples were thawed on ice. Additionally, the 1 mm electroporation cuvettes were kept on ice. 33 µL of cells were transferred to the electroporation cuvette, removing any bubbles by gently tapping, drying excess external humidity and placing the cuvette into the electroporator chamber. For the samples treated with YO-PRO-1, 1.5 mL Eppendorf tubes were kept on ice before starting. Samples were then pipetted into the Eppendorf tubes by duplicate, adding 1 µL of 1 mM YO-PRO-1 to the treatment and 1 µL $H_2O$ to the mock, pipetting the volume to mix.

*Basic Equation and Algorithm*

From a semi-microscopic point of view, the movement of an $i^{th}$ ion with mass $m_i$ suspended in water with velocity $v_i$ during cellular electroporation follows Newton's equations of motion and is best described for long simulation periods by a modified Langevin eqn. [37]:

$$\vec{F}_{R,i} - \vec{F}_{r,i} + \vec{F}_{E,i} + \vec{F}_{mem,i} + \vec{F}_{C,i} + \vec{F}_{sr,i} = m_i \frac{d\vec{v_i}}{dt} , \qquad (1)$$

where $\vec{F}_{R,i}$ represents an stochastic force arising from the interaction of ions with water molecules, $F_{r,i}$ is a friction force of ions with water, $F_{E,i}$ is a force due to an externally applied electric field, the force $F_{mem,i}$ arises from the ion's interaction with the membrane and pores, $F_{C,i}$ is an interionic Coulomb force, and $\vec{F}_{sr,i}$ is the short range repulsion force from the overlap of the ions' electron clouds.

The velocity-Verlet algorithm [63] is implemented to keep track of the positions and velocities of the ions with acceleration $\vec{a}_i(t) = \frac{d\vec{v_i}}{dt}$ at a time $t_n$ with time step $\Delta T$, due to its greater efficiency, stability and accuracy over other algorithms such as leap-frog [6], which calculates velocities and positions at different times, and Verlet [67], which does not generate velocities directly. It has been coded in the C++ programming language (gnu-Linux) with objects of a class that contain the mass, charge, velocity, acceleration and position of an ion for a given time interval.

In order to speed up computation time of the forces, besides using Brownian dynamics as opposed to Molecular dynamics, the workload for ions is distributed evenly amongst various core processors using parallel processing. Since the Coulomb, repulsive and image charge forces have a computational time complexity of $O(n^2)$ and they are dependent on other ions' positions, the message passing interface (MPI) protocol is used to pass the ions' positions among processors. In particular, the mpi_allgatherv function is used, which passes messages using an efficient tree algorithm. The speed up curve reaches a plateau at 50 cores, after which computation time for each time step decreases due saturation of the bus with messages. For computing the diffusion coefficient from the Green-Kubo relation (Diffusion coefficient section), ion's velocities are passed as well to the root process using mpi broadcast. For additional improvement in computational time, the simulations are conducted considering finite subregions of the cytoplasm, inner and outer membranes, periplasm and extracellular environment. The program has a modular design and includes a Makefile for compiling.

## Details on the Calculation of Various Forces

### Friction Force (inelastic collision with the surrounding water)

The total friction force acting on an ion from water molecules it hits at a particular instance is opposite to the direction of motion. It is represented by

$$\vec{F_r} = m_i \gamma_i \vec{v}_i(t), \qquad (2)$$

where $1/\gamma_i$ is the relaxation time constant [24][28]. $F_r$ is considered directly proportional to the ion's velocity, as non-linear restoring forces only occur when the ions move extremely fast. The coefficient of friction $m_i\gamma_i$ is given by Stokes law:

$$m_i \gamma_i = 6\pi n_{vis} R_S, \qquad (3)$$

where $R_S$ is the Stokes radius of the hydrated molecule and $n_{vis}$ is the viscosity of liquid water, which as a function of temperature $T$ can be expressed empirically by [33]

$$n_{vis} = A 10^{\frac{B}{T-C}}, \qquad (4)$$

where A = 2.414x10$^{-5}$ N.s/m$^2$, B = 247.8 K , and C = 140 K . Considering the following relation from [66]:

$$\gamma_i = \frac{k_B T}{m_i D_i}, \qquad (5)$$

where $k_B$ is Boltzmann constant and $D_i$ is the bulk diffusion coefficient of an i$^{th}$ ion, the Stokes radius is obtained by substituting Eqn. (5) in (3) to obtain

$$R_S = \frac{k_B T}{6\pi n_{vis} D_i}. \qquad (6)$$

At T = 298 K, $n_{vis}$ = 8.9x10$^{-4}$ N.s/m$^2$, and $D_i$ is given in Table 3.

### Random Force (elastic collision with water molecules)

The interaction of ions with water molecules is represented with isotropic random collision forces which are Markovian in nature with Gaussian distribution. The Gaussian random collision force $F_R$ has a sample space $S_{FR} = (-\infty, \infty)$ [49] and its probability density function (pdf) is given by

$$f(F_R) = \frac{1}{\sqrt{2\pi\sigma^2}} e^{\frac{-|F_R - m_e|^2}{2\sigma^2}}, \qquad (7)$$

where the mean $m_e$ is zero and the standard deviation $\sigma$ is expressed as [24] [66]

$$\sigma = \sqrt{\frac{2 m_i k_B \gamma_i T}{\Delta T}}. \tag{8}$$

Since the pseudo-random floating point numbers $x_i$ generated by a computer algorithm have a sample space $Sx_i = (0,1)$, for a random variable X with uniform pdf, $f(F_R)$ is normalized. Hence, for every $x_i$ there is a corresponding $F_{Ri}$ such that

$$x_i = \frac{\int_0^{F_{Ri}} e^{\frac{-F_R^2}{2\sigma^2}} dF_R}{\int_0^{\infty} e^{\frac{-F_R^2}{2\sigma^2}} dF_R}. \tag{9}$$

Peforming the change of variable $t = F_R/\sqrt{2}\sigma$ and multiplying both sides of Eqn. (9) by $2/\sqrt{\pi}$ :

$$x_i = \frac{\frac{2}{\sqrt{\pi}}\int_0^t e^{-t^2} dt}{\frac{2}{\sqrt{\pi}}\int_0^{\infty} e^{-t^2} dt}. \tag{10}$$

The numerator is the Laplace error function erf(t) for $t \geq 0$. The integral in the denominator is $\sqrt{\pi}/2$. Hence erf($\infty$) = 1 and, considering the aproximation in [25], Eqn. (10) becomes

$$x_i = \text{erf}(t) = 1 - [a_i z + a_2 z^2 + a_3 z^3] e^{-t^2}, \tag{11}$$

where the coefficients $a_1$=0.3480242, $a_2$=0.0958798, $a_3$=0.7478556, and $z = 1/(1 + pt)$ with p=0.47047.

To save computational time, a look-up table implemented as a 2D array is used to store 400 error function values between 0 and 1 in one column, which are matched to the generated pseudo-random numbers, and the corresponding Gaussian random forces in the other column.

*Coulomb Force (other ions in the reservoir)*

The Coulomb force considered for the n-ion system is given by

$$\vec{F}_{Ci} = \frac{q_i}{4\pi\varepsilon_0} \sum_{\substack{k=0 \\ k \neq i}}^{n} \frac{q_k}{|\vec{r}_i - \vec{r}_k|^2}, \tag{12}$$

where $\vec{F}_{Ci}$ is the net force on an $i^{th}$ ion due to all other ions within the system, $\vec{r}_i$ and $\vec{r}_k$ are the position vectors of the ions and $q$ is the ion electric charge. For simulation purposes, the vector $\vec{F}_{Ci}$ is broken down into its Cartesian projections. For the x axis it is given by

$$F_{Ci,x} = \frac{q_i}{4\pi\varepsilon_0} \sum_{\substack{k=0 \\ k \neq i}}^{n} \frac{q_k(r_{i,x}-r_{k,x})}{\left(|r_{i,x}-r_{k,x}|^2+|r_{i,y}-r_{k,y}|^2+|r_{i,z}-r_{k,z}|^2\right)^{3/2}}. \quad (13)$$

$\vec{F}_{Ci,y}$ and $\vec{F}_{Ci,z}$ have similar expressions to Eqn. (13), except that $r_{i,x} - r_{k,x}$ in the numerator is replaced by $r_{i,y} - r_{k,y}$ or $r_{i,z} - r_{k,z}$, respectively.

### Image Charge Force (charges induced at the membrane)

Considering the effects arising from the difference in dielectric properties between the membranes and surrounding water, due to dielectric polarization when an ion gets close to a non-conducting membrane wall, it induces bound charges within the membrane's surface, which in turn induce a force $\vec{F}_{mem,i}$ on the ion itself. For simple geometries, this phenomenon can best be treated by the method of image charges [59].

The membrane surfaces outside and inside the cell in the area surrounding the pore are approximated with planar boundaries, so that $\vec{F}_{mem,i}$ is taken into account by a Coulomb force (Eqn. 12) due to an image charge $q'$ [59]:

$$q'_i = q_i \left(\frac{\varepsilon_w - \varepsilon_m}{\varepsilon_w + \varepsilon_m}\right), \quad (14)$$

where $\varepsilon_m$ and $\varepsilon_w$ are the permittivities of the membranes and the water solvent, respectively. In terms of $\varepsilon_0$, $\varepsilon_m = K_m \varepsilon_0$ and $\varepsilon_w = K_w \varepsilon_0$ where $K_m$ and $K_w$ are the relative permittivities of the membranes and water.

The charges induced close to the surface of the pore formed by electroporation are approximated considering the Green function of the interior of a cylinder of radius $\rho_0$ and length $2z_0$ [64][65].

The z component on a charge q can be found by evaluating Green's function $G$ at the cylindrical coordinates $\rho' = \rho, \emptyset' = \emptyset, z' = z$ and taking the gradient following Smythe [60]:

$$F_z = -\frac{q^2 \nabla G(\rho,\emptyset,z,\rho',\emptyset',z')}{2} \quad (15)$$

The radial force $F_\rho$ is calculated considering the charge distribution σ on the cylinder surface S, which can be found from an alternate form of Green's function [70]:

$$\sigma = -\varepsilon_0 q \nabla G(\rho, \emptyset, z, \rho', \emptyset', z'), \quad (16)$$

where the gradient is taken with respect to $\rho, \emptyset, z$. Instead of using Coulomb's law to calculate the force on the ion due to σ, it is easier to calculate the total force on the conducting membrane pore, which is the negative of the force on the ion (Newton's third law).

Thus, the radial force on the ion is obtained from

$$F_\rho = \int_S \frac{\sigma^2 da}{2\varepsilon_0}. \tag{17}$$

The total $\vec{F}_{mem,i}$ acting on an $i^{th}$ ion is the vector addition of forces with components $F_z$ and $F_\rho$ arising from its own induced image charge as well as all other image charges induced by other ions in the pore (Details of derivations provided in the Appendix).

### Force from Externally Applied Electric Field

The Lorentz force acting on an $i^{th}$ ion is given by

$$\vec{F}_{E,i} = q_i(\vec{E} + v_i \times \vec{B}), \tag{18}$$

where $\vec{E}$ and $\vec{B}$ are the electric and magnetic fields acting on the system. $\vec{E}$ has two components: one due to the externally applied pulse throughout the cuvette, as explained in the experimental setup section, and another one at the membrane related to the resting potential arising from differences in intracellular and extracellular ionic concentrations. The resting potentials $V_{rest}$ at the inner or outer membranes of *E. coli* (Table 2) enhanced by the externally induced voltage $V_m$ are related to the $\vec{E}$ field as follows:

$$V_{rest} + V_m = V_a - V_b = -\int_b^a \vec{E} d\vec{l}, \tag{19}$$

where points a and b are at the edges of the pores as shown in Figure 2, and $d\vec{l}$ is an infinitesimally small section of its total length. In the simulated system, $\vec{E}$ is assumed to remain constant in space over the entire region of integration (i.e. the pore's area) for each time step, while exponentially decaying over consecutive time steps due to the nature of the applied pulse. The magnetic field $\vec{B}$ is induced by the change in the electric field $\frac{dE}{dt}$ due to the discharge of the capacitor, with magnitude

$$B = \mu_0 r_p I / 2 A_e, \tag{20}$$

where μ₀ is the is the permeability of free space, $r_p = \sqrt{x^2 + y^2}$ is the position of the ions in a plane parallel to the electrodes with area $A_e$ = 200 mm², and $I = V_0 e^{-t/RC}/R$ is the current flowing from the capacitor through the wires connected to the electrodes.

The induced transmembrane potential in relation to the shape and size of the long prolate spheroidal *E. coli* cell and the externally applied electric field $\vec{E}$ is given by [19]

$$V_m = fgr_l E\cos(\theta), \tag{21}$$

where the form factor $f = 1.12$ if the long axis of *E. coli* ($L_{axis}$ in Table 2) and the field are parallel (angle $\theta = 0$), and $f = 1.8$ if they are perpendicular [19], the physiological factor $g = 1$ assuming the membrane is a pure dielectric, and $r_l$ is half the axis of the cell in the direction of the field, The induced potential is highest when the rod is parallel to the field since the shorter axes are three times smaller than the long axis (1:1:3 axial ratio). To calculate $V_m$ at the inner membrane, the periplasm and outer membrane widths are subtracted from the cell's axis.

Thus, when ions are in the pores, the electric field in Eqn. (18) is calculated by substituting $V_m$ from Eqn. (21) and $V_{rest}$ from Table 2 into Eqn. (19), divided by the membranes' thickness. This approach is in accordance with standard *RC* cell theory for small electric fields [15]. However, experiments showed a flattening of $V_m$ to approximately 1V in the polar regions for larger $\vec{E}$ fields [15]. Thus, a cutoff has been implemented for voltages above 1V. The $\vec{E}$ field is different at the inner and outer membranes, since the membranes' thickness, resting and induced potentials are different. In regions surrounding the membrane, $\vec{E}$ in Eqn. (18) is the externally applied electric field. The induced magnetic field $\vec{B}$ is calculated using Eqn. (20) throughout the simulation regions, since only the external $\vec{E}$ field component changes with time, assuming $V_{rest}$ remains constant.

*Short Range Repulsion Force from the Overlap of the Ions' Electron Clouds*

For ions having different charges with opposite polarities, a short range repulsion force arising from the overlap of the electron clouds is considered. In this regard, an interatomic potential U, based on the Thomas-Fermi-Dirac approximation can be expressed in the so-called Born-Mayer form within an interval of internuclear separation R [1] [10]:

$$U(R) = A_s e^{-bR}, R_l \leq R \leq R_{ii}, \tag{22}$$

where $R_l$ and $R_{ii}$ are the lower and upper limits of the interval, typically ~$1.5a_0$ ($a_0 = 0.52917$ Å) and ~$3.5a_0$, respectively, and parameters $A_s$ and b are constants for a given pair of same-type atoms. For pairs of unlike atoms 1 and 2, the following empirical "combination rule" is used:

$$U_{12} \approx (U_{11}U_{22})^{1/2} \approx \left(A_{s1}A_{s2}e^{-(b_1+b_2)R}\right)^{1/2}. \tag{23}$$

The interatomic force is $\vec{F}_{sr,i} = -\nabla U_{12}$.

## Pore Creation and Evolution

Under a typical physical scenario, pores are formed in the cell membrane during electroporation, reach a maximum radius, and subsequently lead to plasma membrane recovery through shrinkage, or membrane disintegration. The rate of pore creation and destruction is governed by the induced potential $V_m$ (Eqn. 21) according to the following first order ordinary differential equation [15][46]:

$$\frac{dN}{dt} = \alpha e^{(V_m/V_{ep})^2} \left(1 - \frac{N}{N_0} e^{-\beta(V_m/V_{ep})^2}\right), \quad (24)$$

where $N$ is the pore density, and $α$, $β$, $V_{ep}$, and $N_0$ are constants provided in Table 2. The total number of pores $K$ at a given time is the surface integral of the pore density: $K = \iint_{membrane} N dA$. Since pores concentrate on both ends of the *E. coli* rod after it rotates parallel to the field (angle θ goes to zero in $V_m$), their surface areas combined can be approximated as $A=2\pi r_w^2$, where $r_w$ is half the shortest axis ($L_{axis,short}$ in Table 2), i.e. $K=2\pi r_w^2 N$. Eqn. (24) is solved with the initial condition $N = 0$ (no pores).

Due to the differences in transmembrane voltage, thickness, position and composition of both membranes, the sizes of the inner and outer pores evolve differently over time to minimize the energy of the lipid bilayers. The change in the radius $r_i$ of an $i^{th}$ pore is determined by the discretized equation

$$\frac{dr_i}{dt} = \frac{D_p}{k_B T} \left\{\frac{F_{max}V_m^2}{1+r_h/r_i} + 4C\left(\frac{r_*}{r_i}\right)^4 \frac{1}{r_i} - 2\pi\gamma - 2\pi\sigma_{eff} r_i\right\}. \quad (25)$$

The first term accounts for the electric force induced by the local transmembrane potential $V_m$ and applies to pores with arbitrary size and a cylindrical inner surface [48]; the second term accounts for the steric repulsion of lipid heads [35][46]; the third, for the line tension acting on the pore perimeter [35][47][46]; and the fourth, for the surface tension of the cell membrane, where the effective tension $\sigma_{eff}$ is given by [47][46]

$$\sigma_{eff} = 2\sigma' - \frac{2\sigma' - \sigma_0}{(1-A_p/A)^2}, \quad (26)$$

where $\sigma_0$ is the tension of the membrane without pores, $\sigma'$ is the energy per area of the hydrocarbon-water interface, and $A_p = \sum_{i=1}^{K} \pi r_i^2$ is the combined area of pores at a given time [47][35], which are initially created with minimum radius $r_i = r_*$ [35]. The average pore radius is $r_{avg} = \sum_{i=1}^{K} r_i / K$. Parameters in Eqns. (24), (25) and (26) are defined in Table 2, assumed to be the same for both inner and outer membranes except $V_m$, which has the greatest effect.

## Model Variables, Physical and Chemical Parameters

Table 1 lists the model variables which are tuned to maximize ion throughput while keeping electroporation reversible. Table 2 lists the physical, chemical, geometry and electroporation constants used in the calculations of the theoretical Brownian dynamics model. Table 3 lists the physical parameters for the ions considered in the model.

**TABLE 1**

**Model variables for modeling experimental conditions.**

| VARIABLE | DESCRIPTION | UNITS |
|---|---|---|
| $T$ | Temperature | K |
| $N_{\Delta T}$ | Number of time steps $\Delta T$ | a.u. |
| $N_{ions}$ | Number of ions | a.u. |
| $E_x, E_y, E_z$ | Applied Electric field projections over Cartesian axes | V/m |
| $\tau$ | Pulse duration | s |
| $r$ | Pore radius | m |

**TABLE 2**

**Physical, chemical, geometry and electroporation parameters used in the theoretical model**

| PARAMETER | DESCRIPTION | VALUE | UNITS | SOURCE |
|---|---|---|---|---|
| $e$ | electron charge | $1.60218 \times 10^{-19}$ | C | 12 71 |
| $m_e$ | electron mass | $9.1094 \times 10^{-31}$ | Kg | 71 |
| $k_B$ | Boltzmann constant | $1.381 \times 10^{-23}$ | J/K | 71 |
| $\varepsilon_0$ | permittivity of free space | $8.8542 \times 10^{-12}$ | $C^2/(N.m^2)$ | 71 |
| $\mu_0$ | permeability of free space | $1.2566 \times 10^{-6}$ | $N/A^2$ | 71 |
| $K_w$ | relative permittivity of water | 80.4 | a.u. | 71 |
| $K_m$ | relative permittivity of membranes | 2 | a.u. | 12 |
| $V_{rest,inner}$ | resting potential – inner membrane | -0.140 | V | 41 |
| $V_{rest,outer}$ | resting potential – outer membrane | -0.050 | V | 2 |
| $\Delta T$ | simulation time constant | $2 \times 10^{-15}$ | s | 12 |
| $L_{inner}$ | inner membrane width | $5 \times 10^{-9}$ | m | 18 |
| $L_{outer}$ | outer membrane width | $13 \times 10^{-9}$ | m | 4 |
| $L_{periplasm}$ | periplasm region width | $11 \times 10^{-9}$ | m | 22 |
| $L_{axis, long}$ | *E. coli* longer axis | $1.5 \times 10^{-6}$ | m | 53 |
| $L_{axis, short}$ | *E. coli* shorter axis | $0.5 \times 10^{-6}$ | m | 53 |
| $Vol_{EColi}$ | *E. coli* volume | $1.571 \times 10^{-18}$ | $m^3$ | 53 |
| $\alpha$ | Pore creation rate coefficient | $1 \times 10^9$ | $m^{-2} s^{-1}$ | 15 46 |
| $\beta$ | Electroporation constant | 2.46 | a. u. | 15 |
| $V_{ep}$ | Characteristic voltage of electroporation | 0.258 | V | 15 46 |
| $N_0$ | Equilibrium pore density at $V_m = 0$ | $1.5 \times 10^9$ | $m^{-2}$ | 15 46 |
| $D_p$ | Diffusion coefficient for pore radius | $5 \times 10^{-14}$ | $m^2 s^{-1}$ | 35 46 |
| $F_{max}$ | Lipid bilayer max electric force for $V_m = 1V$ | $0.638 \times 10^{-9}$ | $N V^{-2}$ | 48 |
| $r_h$ | Constant for advection velocity | $1.466 \times 10^{-9}$ | m | 48 |
| $r_*$ | Minimum radius of hydrophilic pores | $0.51 \times 10^{-9}$ | m | 35 46 |
| $C$ | Lipid bilayer steric repulsion energy | $1.4 \times 10^{-19}$ | J | 35 46 41 |
| $\gamma$ | Lipid bilayer edge energy | $1.8 \times 10^{-11}$ | $J m^{-1}$ | 35 46 47 |
| $\sigma'$ | Tension of hydrocarbon water interface | $2 \times 10^{-2}$ | $J m^{-2}$ | 47 |
| $\sigma_0$ | Tension of the bilayer without pores | $1 \times 10^{-6}$ | $J m^{-2}$ | 35 46 |

# TABLE 3

**Physical parameters for various ions**

|  | $Ca^{2+}$ | $Mg^{2+}$ | $K^+$ | $Na^+$ | $Cl^-$ |
|---|---|---|---|---|---|
| $m_i$ (kg) mass | 6.6x10$^{-26}$ [a] | 4.04x10$^{-26}$ [b] | 6.5 x10$^{-26}$ [c] | 3.8x10$^{-26}$ [a] | 5.9x10$^{-26}$ [a] |
| $q_i$ (C) charge | 2e | 2e | e | e | -e |
| $D_i$ (m²/s) bulk diffusion (298 K) | 0.79x10$^{-9}$ [a] | 0.71x10$^{-9}$ [b] | 1.96x10$^{-9}$ [c] | 1.330x10$^{-9}$ [a] | 2.03x10$^{-9}$ [a] |
| $A_s$ (eV) in Eqn.(22) | 8124.1[d] | 3829[d] | 7563[d] | 3661.4[d] | 6411.8[d] |
| b (Å) in Eqn. (22) | 3.61026[d] | 3.69813[d] | 3.62141[d] | 3.77759[d] | 3.63681[d] |
| Intracellular concentration in *E coli* (mM) | 3[e] | 100[e] | 300[e] | 10[e] | 200[e] |
| Extracellular concentration (mM) | 1[e] | 0.5[e] | 5.4[e, f] | 140[e, f] | 151[e,f] |

[a]Reference 13
[b]Reference 40
[c]Reference 14
[d]Reference 1
[e]Reference 45
[f]Reference 7

## Initial Conditions Given to the Simulated Ions

The Maxwell-Boltzmann probability density function $f(v)$ describing the distribution of molecular speeds in terms of the translational kinetic energy $\varepsilon_i$ of a an i$^{th}$ ion is given by

$$f(v) = \frac{8\pi}{m_i}\left(\frac{m_i}{2\pi k_B T}\right)^{\frac{3}{2}} \varepsilon_i e^{\frac{-\varepsilon_i}{k_B T}}, \tag{27}$$

Since the pseudo-random floating-point numbers $x_i$ has a sample space $Sx_i$ = (0,1), for a random variable x with uniform pdf, f($F_R$) is normalized. Hence, for every $x_i$ there is a corresponding $\varepsilon_i$ such that

$$x_i = \frac{\int_0^{\varepsilon_i} e^{\frac{-\varepsilon_i}{k_B T}} d\varepsilon_i}{\int_0^{\infty} e^{\frac{-\varepsilon_i}{k_B T}} d\varepsilon_i}. \tag{28}$$

The integrals in the above equation are solved directly giving

$$x_i' = e^{\frac{-\varepsilon_i}{k_B T}}, \tag{29}$$

where $x_i' = 1 - x_i$, it follows

$$\varepsilon_i = -k_B T \ln(x_i') . \tag{30}$$

Substituting the kinetic energy Eqn. $\varepsilon_i = m_i v_i^2/2$ in the above, we get the initial speed of an i<sup>th</sup> ion:

$$v_i = \sqrt{\frac{-2k_B T \ln(x_i')}{m_i}}, \tag{31}$$

noting that ln(x') is a negative number. The ion's initial velocity vector with magnitude equal to its initial speed is in a direction specified by the spherical angles $\theta$ and $\varphi$, which are randomly calculated based on a uniform pdf. The velocity spherical coordinates are transformed to Cartesian when incorporated in the model.

The initial positions of ions have been randomly assigned with uniform pdf within the cytoplasm subregion under simulation.

## Geometry and Boundary Conditions of the Simulated Regions

The *E. coli* rod is subjected to a torque and rotates because of the induced dipole, reorienting itself with the longest axis parallel to the applied field direction. Only at the end of orientation permeabilization can be efficient [19]. In the simulation, pore formation is considered in the inner and outer membranes (Figure 2). Both pores are aligned with each other in a membrane region at the hemispherical end of the rod along its larger axis in the z direction parallel to the field. In this position, the highest transmembrane potential is induced (Eqn. 21), and thus there is a higher probability of pore formation.

For practical purposes, the pores are modeled with a cylindrical shape with sharp boundaries and evolve over time according to Eqns. (24) to (26). The inner and outer sides of the cell are modeled with finite cubic volumes with imaginary or soft boundaries (See Figure 2).

When an ion leaves the system, another ion is symmetrically inserted back from the opposite boundary, thereby keeping the number of particles in the system constant in order to approximate a canonical ensemble in thermal equilibrium with a fixed temperature and volume. For the sake of simplicity and computational efficiency, this approach has been chosen over stochastic Monte Carlo techniques [31], which are more complex but do not necessarily improve the accuracy of the predictions [14]. The initial specific ion concentrations in the intracellular and extracellular subregions are provided in Table 3. Ions leaving the extracellular subregion across the boundary opposite the outer membrane are reinserted in the cytoplasm.

In order to avoid short range interactions when applying periodic boundary conditions, the shortest side of the cytosol, periplasm and extracellular boxes have been chosen to be at least twice as big as the cut-off radius for short range electrostatic forces [16]. The box sizes are sufficiently small to keep computational efficiency while having a proportional ionic concentration, with 10 nm sides for 1 and 4 nm pore radius, and 25 nm to contain the pores with 20 nm radius. Bigger simulation boxes have a negligible effect on both the thermodynamics and kinematics of ions at the studied regime [20]. On the other hand, water diffusion in MD simulations is affected by the box size [20]. In the case of the present BD work, however, water is treated as a continuum and its diffusion is hence unaffected.

Interactions between the ions and the pore or the cell membrane surrounding it are elastic in nature, i.e., the total momentum and kinetic energy are conserved, and the collisions are handled with specular scattering at hard boundaries, with the angle of incidence equal to the angle of reflection/deflection.

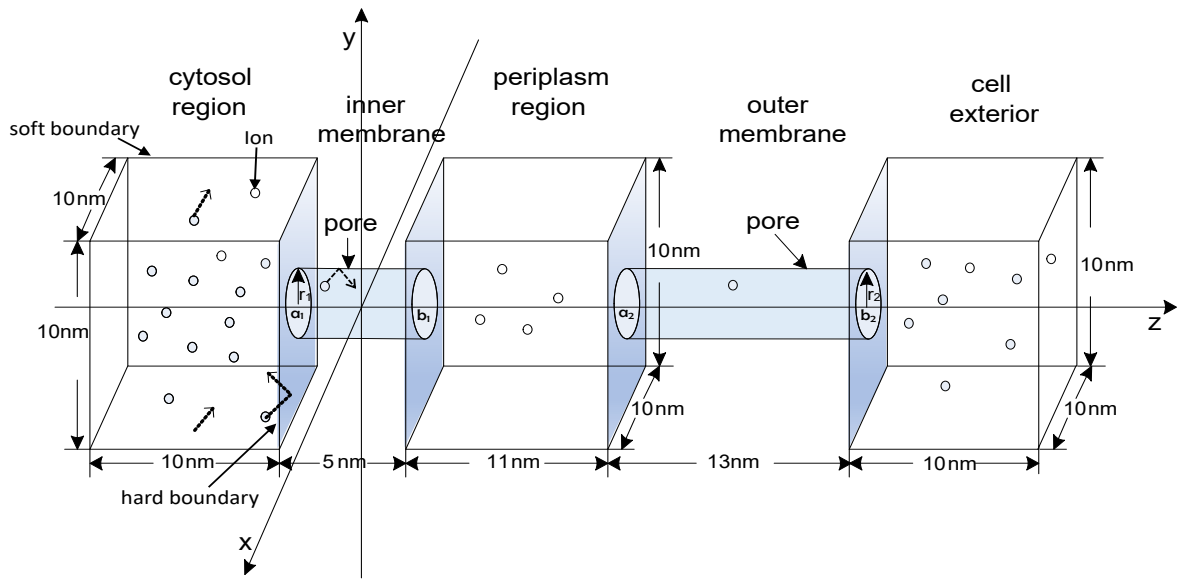

**FIGURE 2.** Geometry and dimensions of the simulated region of an electroporated *E. coli* bacteria, containing the cytoplasmic and external membranes. $a_1$, $b_1$ are the edges on the z axis of the inner membrane pore with radius $r_1$; $a_2$, $b_2$ are the edges of the outer membrane pore with radius $r_2$. Ions are pushed through the pores by the electric field towards the periplasm between the membranes and the extracellular region. Hard boundaries are highlighted in blue. Other boundaries are soft in the cytosol or opened in the periplasm and extracellular region. (Dimensions as in Table 2).

## Calculation of Transport Parameters

### Diffusion Coefficient

The diffusion coefficient *D* is calculated by integrating the velocity autocorrelation function *C(t)* for 3 dimensions (Green-Kubo relation) [23] [36]:

$$D = \frac{1}{3}\int_0^\infty C(t)dt .\quad (32)$$

*C(t)* is given by

$$C(t) = <v_i(t)v_i(t_0)>,\quad (33)$$

where $v_i(t_0)$ is the ion's initial velocity and the notation <> signifies the average. When evaluated over *N* ions, the above Eqn. is expressed as follows:

$$C(t) = \frac{1}{N}\sum_{i=1}^{N}[v_i(t)v_i(t_0)].\quad (34)$$

The integration of C(t) is approximated with the trapezoidal rule.

### Mobility

The diffusion coefficient D of ions is related to their mobility by Einstein's relation [17]:

$$\frac{D}{\mu} = \frac{k_B T}{q}.\quad (35)$$

### Translation (Membrane crossing) Time

Typically, only a few ions flow through a pore simultaneously because of the small pore size and repulsive forces among ions. Thus, simulations were run for only one ion crossing the pore at a time, thereby saving computational time (approximately seven hours for each ion crossing on one 2.1 Ghz core, different crossings were sent to different cores with MPI in parallel). In the case of bigger pores with a 20 nm radius or greater, there is a possibility of multiple ions being present simultaneously. Thus, a multi-ion treatment was utilized for calculating the membrane crossing time in bigger pores, as well as for determining the conductivity validated with experimental data.

*Current and Conductivity*

The conductivity for an *n* concentration of ions is given by

$$\sigma = qn\mu. \tag{36}$$

The current in the pore can be calculated by

$$I = \int J dA. \tag{37}$$

The current density *J* is the current per cross sectional area A:

$$J = \frac{I}{A} = \sigma E + Dq\nabla n. \tag{38}$$

The first term on the right-hand side is the drift current (Ohm's law), the second term is the diffusion current (Fick's 1st law)

Also from Eqns (36) to (38), the drift current is

$$I_{dr} = nAvq. \tag{39}$$

**RESULTS**

*Diffusion*

*Calculation of Diffusion Coefficient in Bulk for Model Validation*

The bulk diffusion coefficient is known from experimental observations, and hence it is possible to compare with a computer simulation to verify the accuracy of the calculations. Figure 3 shows the velocity autocorrelation function (Eqn. (33)) as a function of time for 1000 ions of each type suspended in a bulk system. The calculated bulk diffusion coefficients at 298 K (provided in Table 4) are the areas under the curves (Eqn. (32)) which closely match the values reported by other authors (provided in Table 3), with an average root mean square error equal to $2.8132 \times 10^{-10}$ m²/s for ten simulation runs of each ion type. The difference between the values does not have a significant effect on the results during the pulse since the main driving force is due to the applied electric field, as explained later in the discussion section. However, the small difference in the coefficients might have a minor effect on the calculations after the pulse when transport is primarily due to diffusion.

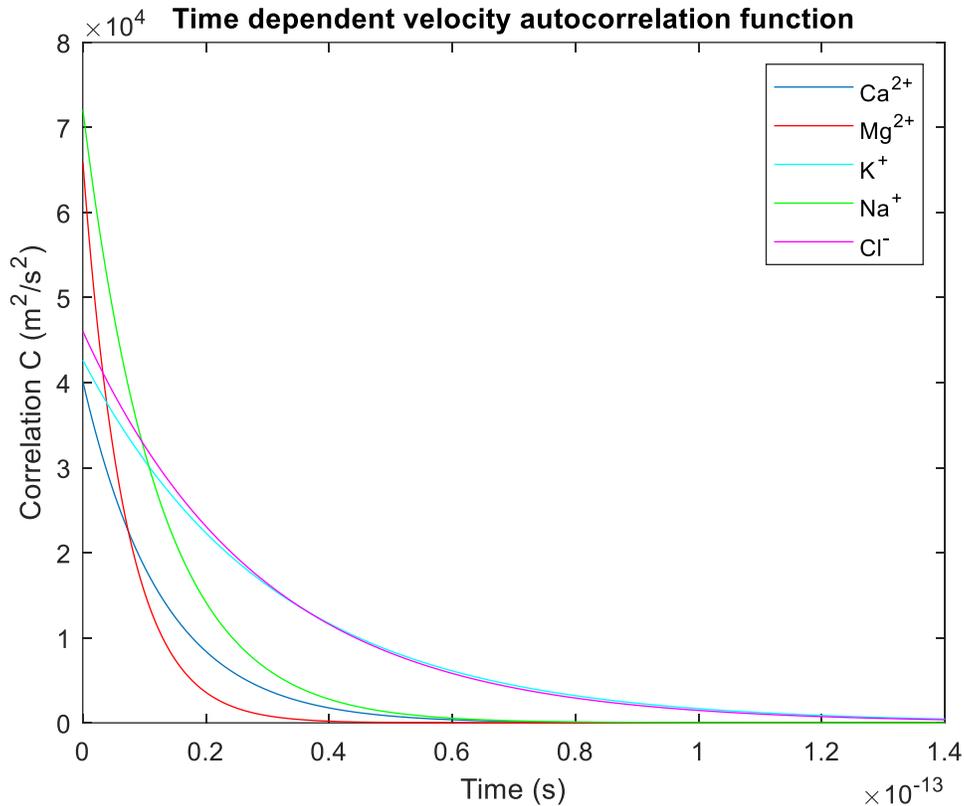

**FIGURE 3**. Time dependent velocity autocorrelation function for 1000 $Ca^{2+}$, $Mg^{2+}$, $K^+$, $Na^+$ and $Cl^-$ ions in a 3D bulk system.

*Diffusion in Pore vs. Temperature*

The dependence of the diffusion coefficient on temperature has also been studied. Simulations were run for ions within the pore under experimental conditions (T = 298 K, E = 18 kV/cm, $r_{avg}$ = 18 nm from Eqn. (25) after 1 ms) and at 278 K and 318 K. These temperatures were chosen so they are not far apart from each other, are within the temperature range for *E. coli* growth (277 K to 322 K – ideally 310 K) and its survival limits, typically greater than T = 193 K (below freezing point) and less than water's boiling point [3]. This temperature range can be easily considered in future experiments. Care is taken to avoid phase change, since at temperatures below 273 K water is frozen so friction increases sharply and Brownian Motion is no longer observed nor can be simulated with the model.

It has been observed that the diffusion coefficient increases as the temperature gets higher, as shown in Figure 4. Error bars represent one standard deviation of uncertainty of ten simulation runs. Smaller ions have less friction with water molecules, thus having higher diffusion with steeper dependence on temperature, which is related to the decrease of water viscosity with heating proportional to the ion radius (Eqns. 2-6) The findings agree with the kinetic theory of gases.

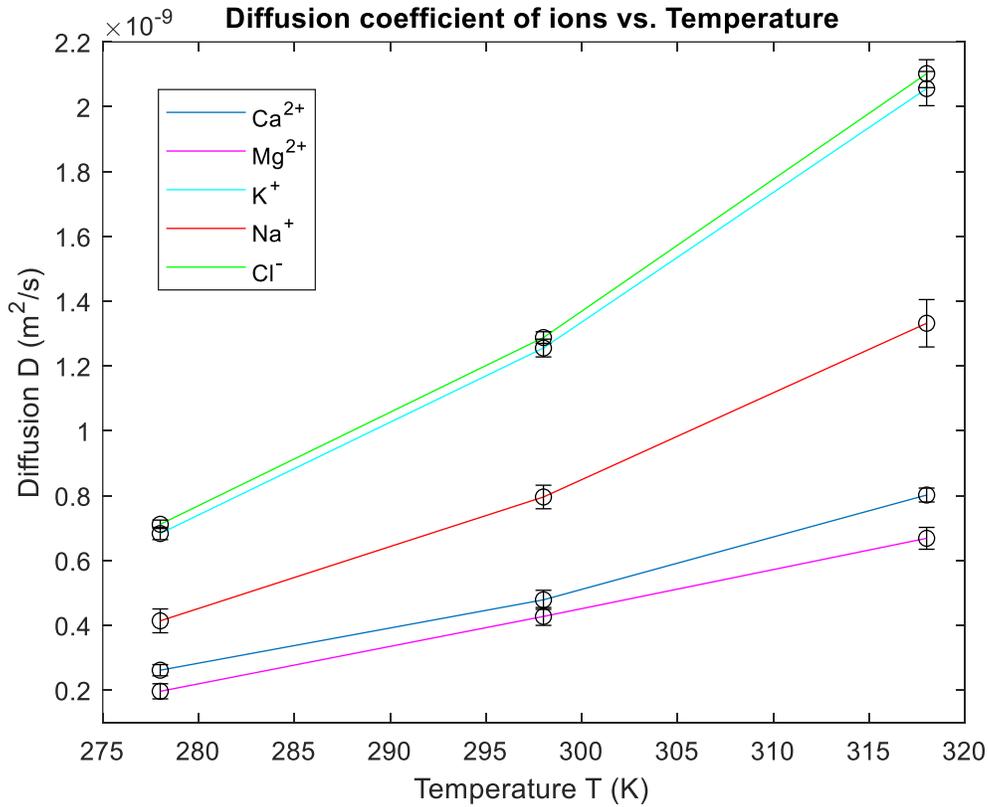

**FIGURE 4**. Dependence of the diffusion coefficient on temperature. Values are shown with standard deviation error bars for $Ca^{2+}$, $Mg^{2+}$, $K^+$, $Na^+$ and $Cl^-$ ions within the pore.

*Diffusion in Pore vs. E-field*

Simulations were run for ions in the pore under the influence of an external force due to an electric field. The results agree with the theory which states that if a field is applied, the particles' diffusion will remain the same because no energy-dependent frictional force or non-uniform scattering process has been incorporated (other than its relation to temperature). In the case of an energy dependent scattering, e.g., the monotonic increase with energy observed in semiconductor transport, the diffusion coefficient would have exhibited a corresponding decrease. Figure 5 depicts the diffusion coefficient values in the pore with one standard deviation error bars of ten simulation runs at the experimental room temperature condition (T

= 298 K) for different magnitudes of an externally applied field. The pore radius is related to the applied electric field by Eqns. (21) and (25). The slight variation of the diffusion coefficient seen in the figure is statistical in origin arising from the Brownian motion of ions due to the surrounding water molecules, accounted for by the random Gaussian force $F_R$.

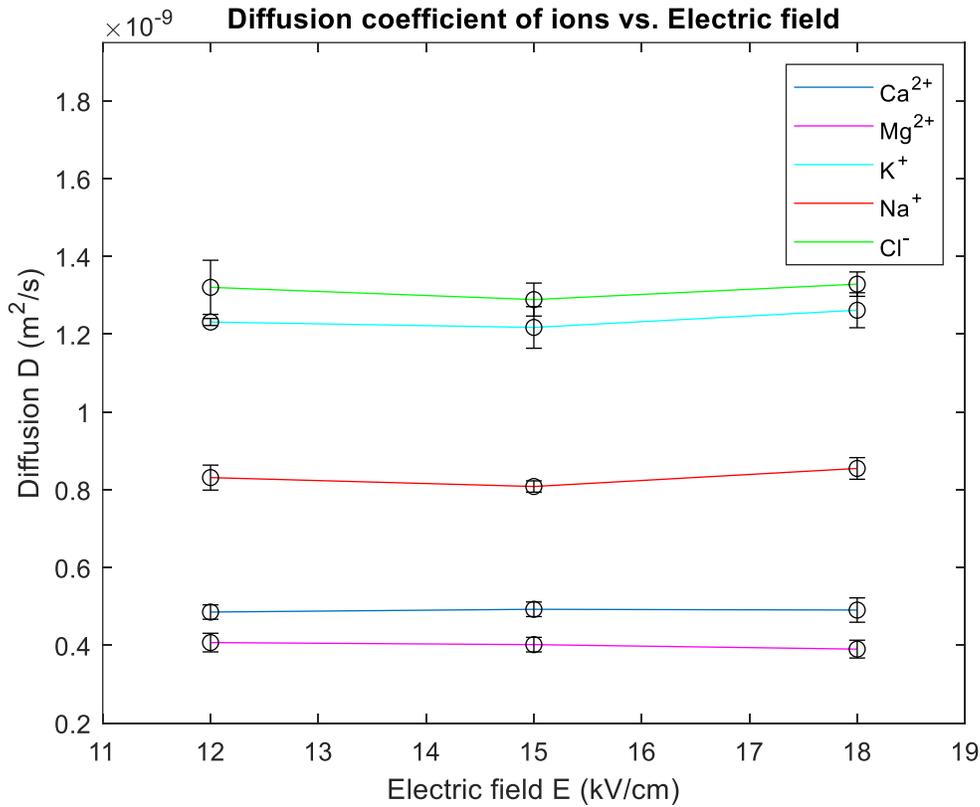

**FIGURE 5.** Diffusion coefficient values with standard deviation error bars for $Ca^{2+}$, $Mg^{2+}$, $K^+$, $Na^+$ and $Cl^-$ ions in the pore for various magnitudes of an externally applied electric field at temperature T = 298 K.

*Diffusion in Pore vs. Pore Size*

The diffusion coefficients of ions in pores with different radius sizes are shown in Figure 6. The effect of small and large pores are explored, with radius values differing by similar factors (x4 or x5). Other conditions correspond to the experiment (T = 298 K, E = 18 kV/cm). It can be concluded from the results in the figure that the diffusion of ions does not vary much according to the pore's radius. This might be due to the fact that random forces do not move the ion far away from the straight trajectory it would otherwise have if no random forces were present. Hence the ion hardly ever touches the boundary, if at all, when initially placed at the center of the pore. Some model results in the literature, however, point at an increase in ionic uptake with increasing pore size, due to a crossover from cation to anion specific permeation [38], or a monotonically decreasing energy barrier with increasing pore radius [32]. In particular, MD simulations found a Cl$^-$ flux increase from 0.02 ns$^{-1}$nm$^{-2}$ for a medium sized pore (Area = 4 nm$^2$) to 0.1 ns$^{-1}$nm$^{-2}$ for a larger pore (Area = 9 nm$^2$), while Na$^+$ showed no significant increase in the permeation rate [38].

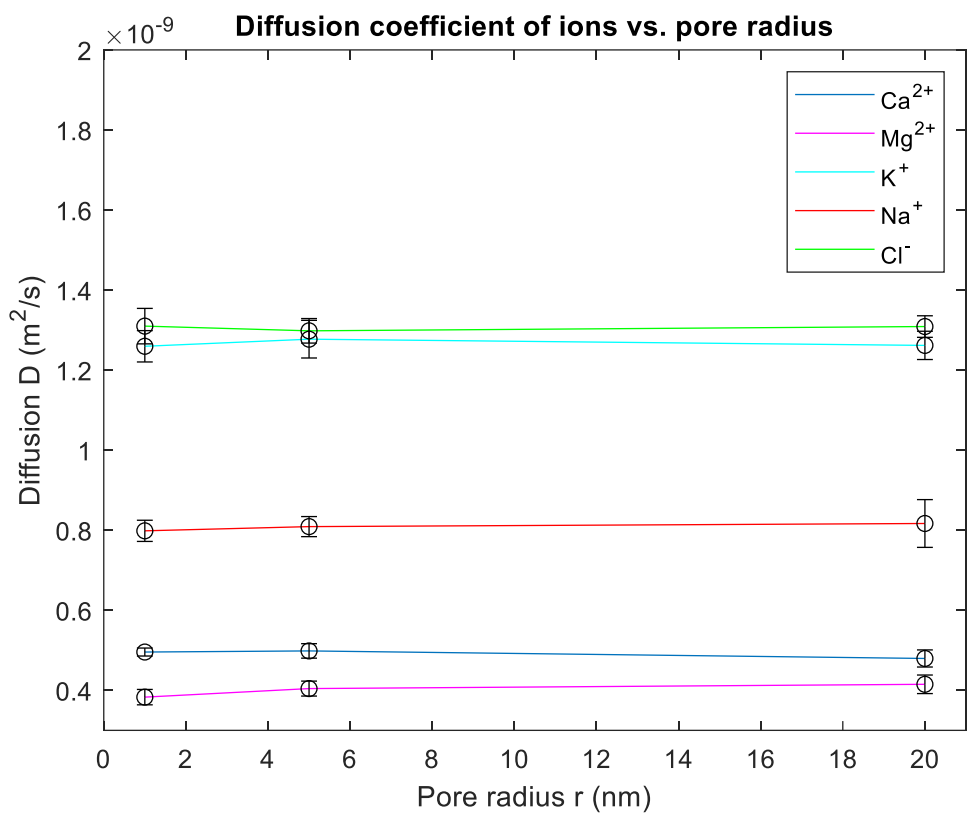

**FIGURE 6.** Diffusion coefficient values with standard deviation error bars for Ca$^{2+}$, Mg$^{2+}$, K$^+$, Na$^+$ and Cl$^-$ ions in the pore for various radius sizes at T = 298 K.

*Mobility and Drift Velocity*

Table 4 below lists calculated mean mobility values of ions moving through the pore using Einstein's Eqn. (35) for the corresponding mean diffusion coefficient of ten simulation runs calculated with the velocity autocorrelation function (Eqns. 32-34) at temperature T = 298 K, applied electric field = 18 kV/cm, pore radius r = 1 nm. The diffusion coefficient values in Table 4 correspond to those shown in Figure 4 at room temperature, and Figures 5 and 6 irrespective of the applied field and pore radius. The drift velocity is the average velocity in the z direction parallel to the pore's axis and is obtained directly from model output.

**Table 4**

**Calculated mean mobility values for the corresponding ionic diffusion.**

|  | $Ca^{2+}$ | $Mg^{2+}$ | $K^+$ | $Na^+$ | $Cl^-$ |
|---|---|---|---|---|---|
| D ($m^2$/s) diffusion | 5.37589x10$^{-10}$ | 4.72354x10$^{-10}$ | 1.19217x10$^{-9}$ | 8.97625x10$^{-10}$ | 1.2181x10$^{-9}$ |
| μ ($m^2$/(V.s)) mobility | 2.09290x10$^{-8}$ | 1.83894x10$^{-8}$ | 4.6413x10$^{-8}$ | 3.49459x10$^{-8}$ | 4.74222ex10$^{-8}$ |

## *Translation (Membrane crossing) time*

### *Histograms of Transport Time of Ions vs. Pore Size*

In order to evaluate ionic throughput dependence on pore size, simulations were run 100 times for each ion type to have sufficient statistics with a low fractional sampling error of 0.142 [8]. Figures 7, 8 and 9 show histograms of $Ca^{2+}$, $Mg^{2+}$, $K^+$, $Na^+$ and $Cl^-$ ions' transport time through pores with radius = 1, 5 and 20 nm under experimental conditions which are standard for bacterial elecgtroporation, i.e., a high 18 kV/cm applied electric field at temperature T = 298 K to permeabilize both the inner and outer membranes of *E. coli*. The $Cl^-$ anion flows in the opposite direction of the cations in pores created on the other side of the cell. Table 5 tabulates the mean and standard deviation of the calculated inner membrane ion crossing times through pores with different radii (shown in Figures 7, 8 and 9). Applying a T-test statistical analysis to these results, it is determined that there is no statistically significant difference in transport time. This might be due to the fact that the random forces arising from water molecules and induced charges at the membrane do not move the ions far away from a straight trajectory propelled by the electric field.

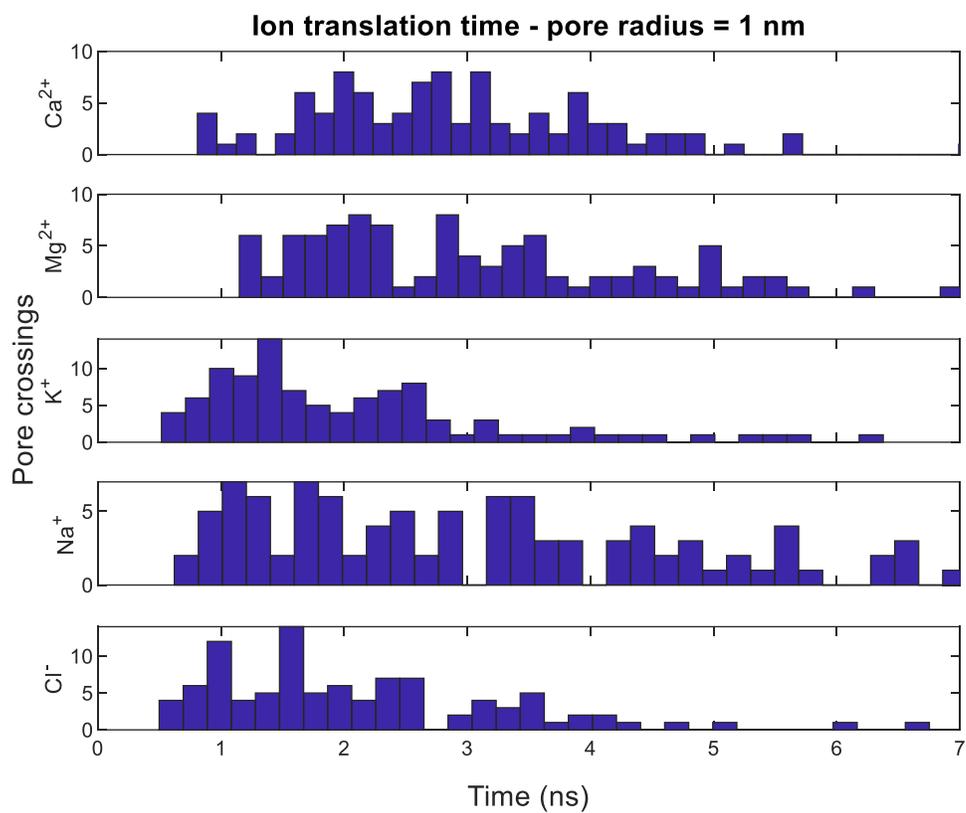

**FIGURE 7.** Histograms of $Ca^{2+}$, $Mg^{2+}$, $K^+$, $Na^+$ and $Cl^-$ ions transport time through a 5nm long pore with radius = 1 nm, under an 18 kV/cm applied electric field at temperature T = 298 K.

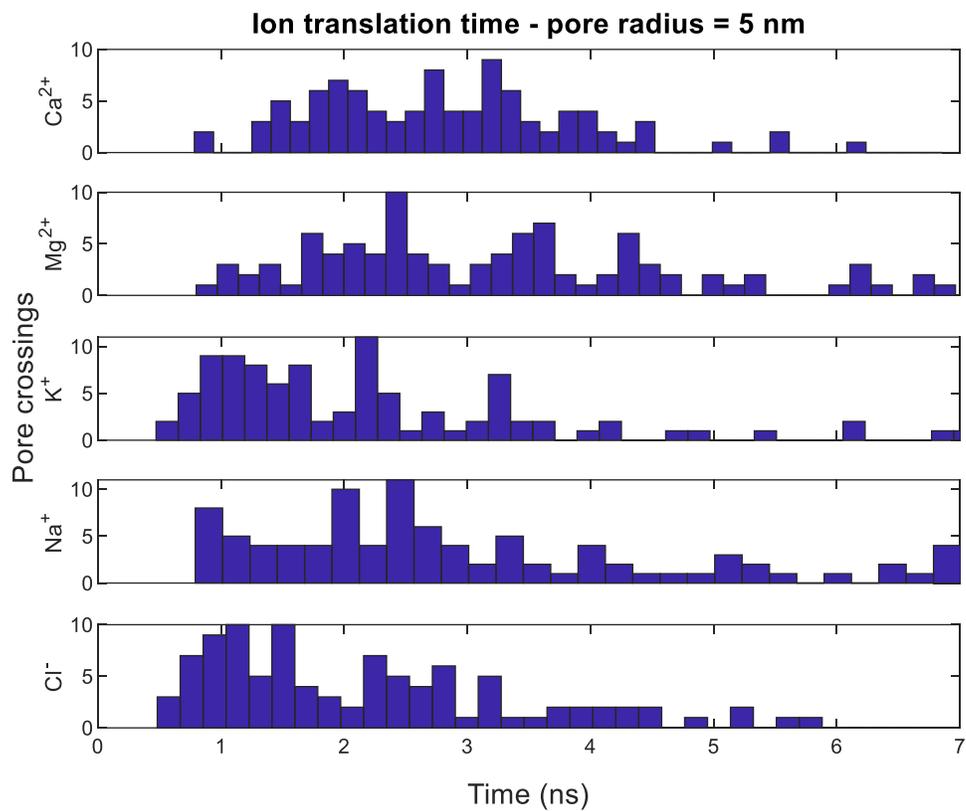

**FIGURE 8.** Histograms of $Ca^{2+}$, $Mg^{2+}$, $K^+$, $Na^+$ and $Cl^-$ ions transport time through a 5nm long pore with radius = 5 nm, under an 18 kV/cm applied electric field at temperature T = 298 K.

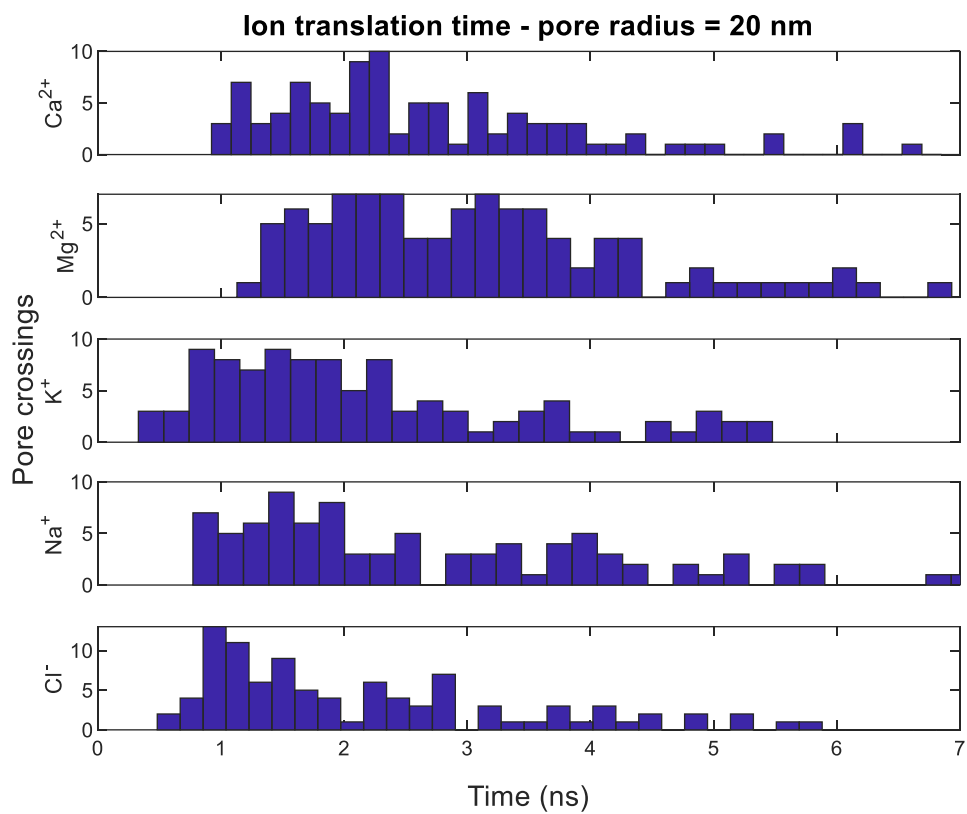

**FIGURE 9.** Histograms of $Ca^{2+}$, $Mg^{2+}$, $K^+$, $Na^+$ and $Cl^-$ ions transport time through a 5nm long pore with radius = 20 nm, under a 18 kV/cm applied electric field at temperature T = 298 K.

**Table 5**

**Mean and standard deviation of calculated inner membrane ion crossing time through pores with various radiuses r**

|  |  | $Ca^{2+}$ | $Mg^{2+}$ | $K^+$ | $Na^+$ | $Cl^-$ |
|---|---|---|---|---|---|---|
| r = 1nm | Mean (ns) | 2.9060 | 3.1709 | 2.0613 | 3.1606 | 2.2750 |
|  | Standard deviation (ns) | 1.1679 | 1.5621 | 1.2134 | 1.8760 | 1.5673 |
| r = 5 nm | Mean (ns) | 2.9802 | 3.4590 | 2.4733 | 3.4058 | 2.3205 |
|  | Standard deviation (ns) | 1.3670 | 1.8574 | 1.8862 | 2.3639 | 1.6061 |
| r = 20 nm | Mean (ns) | 2.7491 | 3.2764 | 2.1897 | 3.6325 | 2.4188 |
|  | Standard deviation (ns) | 1.4137 | 1.6760 | 1.2879 | 3.1238 | 1.7958 |

## Conductivity – Model Validation with Experiments and Comparative Evaluation for Different Pulses

In order to estimate the average number of permeated *E Coli* cells in the samples, measurements were taken with a flow cytometer in reference to an unpulsed control, indicating $1.57 \times 10^{10}$ electroporated cells in the fresh sample, and $1.47 \times 10^{11}$ in the overnight electrocompetent sample. The percentage of electroporated cells is 87.65 % and 85.87 % for the fresh and overnight samples, respectively. Representative graphs of the flow cytometry data are provided in Figure 10.

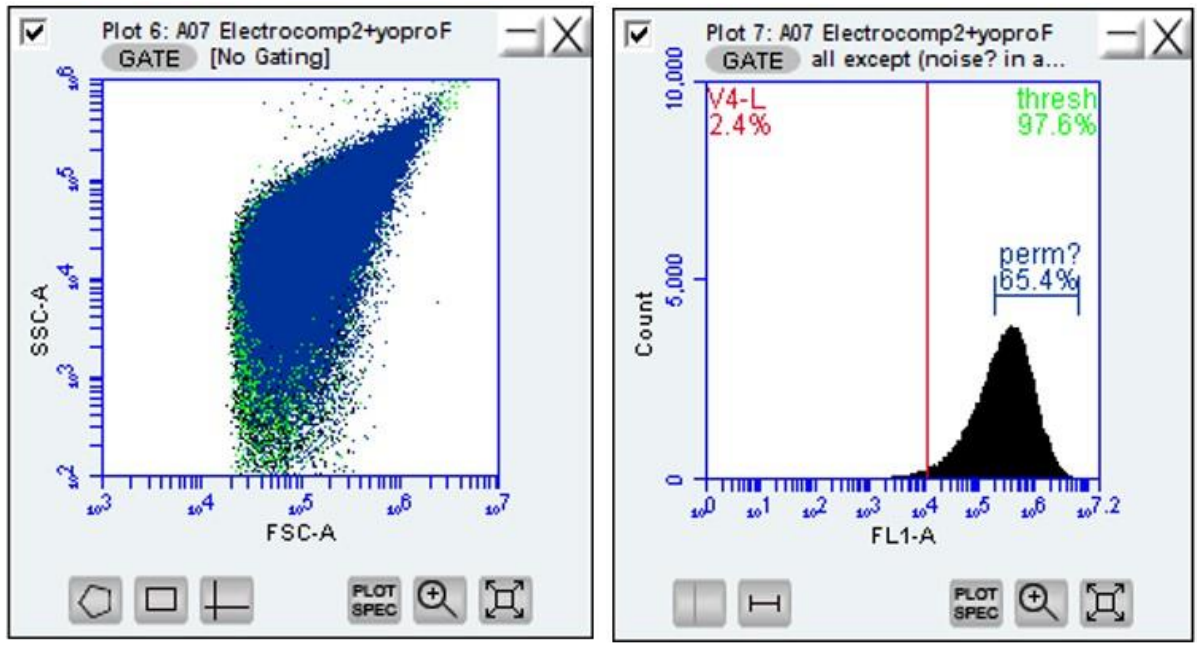

**FIGURE 10**. Flow cytometry graphs of pulsed electrocompetent DH5a *E. coli* cells with YOPRO-1. 33 uL were taken in 200 uL 10% glycerol (in DI Water as blank). Plot 6 depicts the non-electroporated (green) and electroporated w/ YOPRO-1 absorption (blue) populations based on their forward (FSC-A) and side scatter (SSC-A) light intensity profiles. Plot 7 shows a histogram of the permeabilized cell count after gating for various light intensities (FL1-A).

The computed contributions of each ion type to the increase in the conductivity (Eqn. 36) of electrocompetent *E. coli* are shown in Figure 11 as a function of time during the first nanoseconds of the exponential decay pulse with $V_0$ = 1800 V and $\tau$ =7.5 ms. The electric pulse starts 1 ms before the results shown, which is the time it takes for all *E. coli* cells to realign themselves and trigger an efficient permeabilization at their hemispherical ends, with their longer axes parallel to the applied electric field [19]. The conductivities shown in Figure 11 are calculated multiplying the ionic efflux of an inner/outer membrane pore obtained from the BD model times the number of pores per cell considering their average radius, and the total number of cells in the sample from the flow cytometry data. Applying Eqn. (24), 11 inner/outer membrane pores start to be created per cell at time t = 1 ms with an average radius of 35 nm. The discrete jumps in conductivity seen in Figure 11 are related to the nature of the simulation, since ion efflux is stored every 100 time steps. These results are comparable to the temporal growth in ionic concentration depicted in the continuum model work by Q. Hu and R. P Joshi [29], although they focus on calcium ion transport in mammalian cells subjected to monopolar and bipolar square pulses with higher amplitude. Cations flow in the direction of the applied electric field and the anion in the opposite direction in pores created on the other side of the cell. The corresponding drift current density is the first term in Eqn. (38), i.e., the conductivity multiplied by the field.

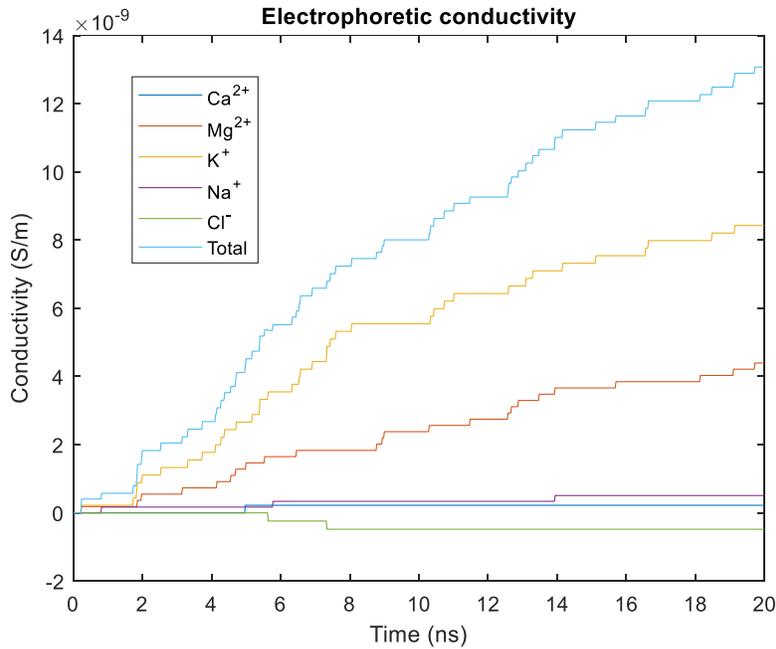

**FIGURE 11**. Computed electrophoretic conductivity of *electrocompetent E. coli* as a function of time for each ion type in the nanosecond regime during the pulse. Time t = 0 corresponds to 11 ms after the pulse rising time.

After application of the pulse, ions continue to diffuse through long-lived pores or permeable structural defects until homeostasis is reached, pores are fully resealed, and the ion membrane permeation barrier is restored. The increase in the reservoir's extracellular concentration of ions $c_{ions}$ with time is characterized quantitatively by the Nernst-Planck equation, which has the following analytical solution for passive diffusion after the pulse [52]:

$$c_{ions}(t) = A + B(1 - e^{Ce^{-\frac{t}{\tau_1}}}), \tag{40}$$

where $A, B$, C and $\tau_1$ have the same meanings as in reference [52]. In the above equation, the exponential decay accounts for the resealing of pores with time. While the BD model simulates a canonical ensemble with a constant number of ions in nanoscale subregions, it keeps track of ions exiting the pores which contribute to an overall increase in the macroscopic reservoir's ionic concentration.

Considering Eqn. (36), the passive diffusion conductivity $\sigma_p$ is related to the ionic concentration as follows:

$$\sigma_p(t) = c_{ions}(t) n_{pores} n_{cells} [0.016 q_{Ca^{2+}} \mu_{Ca^{2+}} + 0.348 q_{Mg^{2+}} \mu_{Mg^{2+}} + 0.579 q_{K^+} \mu_{K^+} \tag{41}$$

$$+ 0.026 q_{Na^+} \mu_{Na^+} + 0.031 q_{Cl^-} \mu_{Cl^-}],$$

where the ionic charges $q$ are provided in Table 3 and the mobilities $\mu$ in Table 4. The equation terms correspond to the conductivity change from individual ions, with the decimal numbers being proportions of the total ion concentration determined by the BD model in nanosecond resolution with no applied field. Under the same external conditions, this proportionality trend remains constant over longer periods of time. The number of pores per cell n_{pores} is calculated with Eqn. (24), n_{cells} is the number of cells in the sample derived from flow cytometry measurements. Thus, it is possible to scale down to an individual pore conductivity for better comparison with the model. During the pulse, the solution to the Nernst-Planck takes the linear form $c_{ions}(t) = kt$ [52].

Figure 12 shows the conductivity values obtained using Eqn. (41) curve fitted using linear regression to the average experimental data of electroporated triplicate samples of fresh and overnight electrocompetent DH5a *E coli* bacteria. The latter sample has a higher conductivity due to its higher permeability and number of cells. Fresh DH5a conductivity measurements were taken every 30 s for 7 mins, starting 45 s after the pulse. Electrocompetent DH5a conductivity measurements were taken every 30 s for 10.5 mins, starting 25 s after the pulse. The average coefficients of variation are 2.3181 % and 4.9292 % for the fresh and overnight triplicate data, respectively. The FiveEasy FiveGo FE30 conductivity probe employed has an experimental uncertainty equal to 0.5 % of the measured values. Considering human reaction time, the time intervals measured with a stopwatch have an absolute experimental uncertainty of 0.2 s. The experimental uncertainty of the time intervals is not depicted with horizontal error bars since it is unnoticeable in the figure's time scale. The model closely fits the measured conductivity data, with root mean square error equal to 2.0839 µS/m for fresh bacteria, and 2.898 µS/m for electrocompetent bacteria. The increase in the conductivity of the suspension over several seconds is in accordance with an experimental report [58] which indicates that longer electric pulses create larger pores which take longer to reseal, e.g., 6-7 mins for a 100 us pulse, 15-20 mins for a 2 ms pulse. In the present study the pulse duration time constant is 7.5 ms.

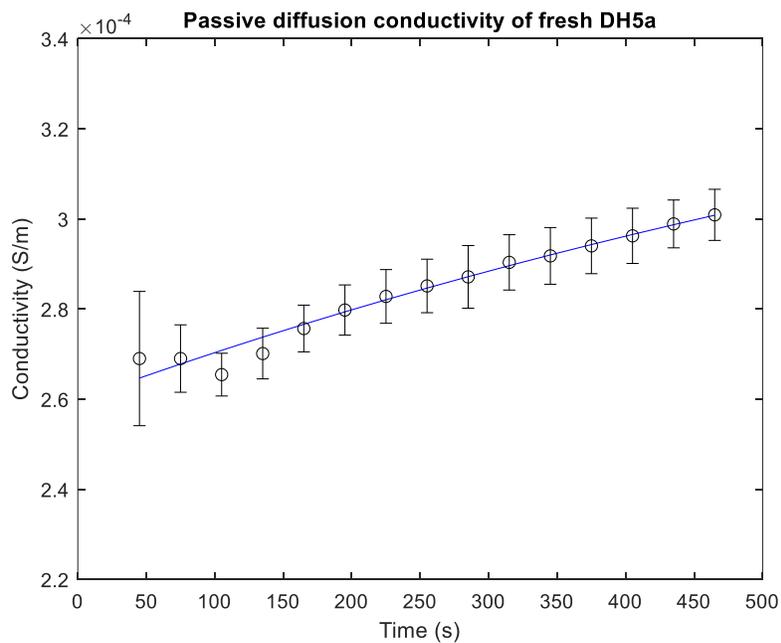

(a)

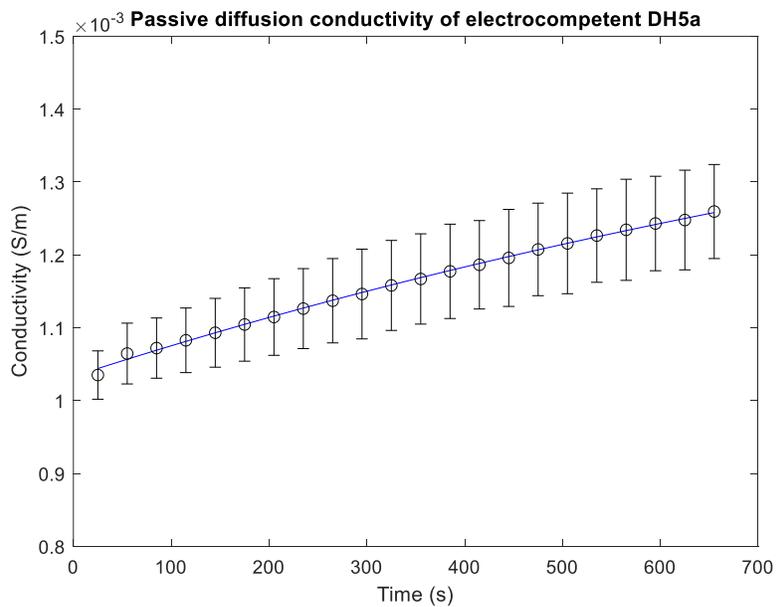

(b)

**FIGURE 12**. Calculated passive diffusion conductivity after the pulse (solid lines) compared to the average experimental data for a) fresh DH5a *E. coli* bacteria and b) electrocompetent DH5a *E coli* grown overnight. Triplicate samples are shown with standard deviation error bars.

According to Eqn. (24), the pore density increases steadily and reaches a maximum value, after which it decreases slowly over several milliseconds. The maximum pore density is related to the magnitude of the applied electric field; the higher the field, the longer it takes to reach a higher maximum value. The average radius decreases over time due to the exponential decay of the electric field and related induced membrane voltage in the first term in Eqn. (25).

We explore the effect on the conductivity of various electric fields ranging from 12 to 21 kV/cm after 11 ms, when pore density has reached a maximum steady value for all fields considered, thus allowing a better comparison. Considering exponential decay pulses, for E(t=0) = 12 kV/cm, 38 pores are formed per cell after 11 ms with average radius $r_{avg}$ = 14 nm; for E(t=0) = 15 kV/cm 47 pores are formed with $r_{avg}$ = 16 nm; for E(t=0) = 18 kV/cm 54 pores are created with $r_{avg}$ = 18 nm; for E(t=0) = 21 kV/cm 60 pores with $r_{avg}$ = 20 nm. Considering a square wave pulse with a constant field E(t) = 18 kV/cm for 11 ns, 58 pores are created with $r_{avg}$ = 20 nm.

Figure 13 shows the conductivity of electrocompetent *E. coli* for different applied electric fields for a nanosecond time period starting 11 ms after the pulse rise time (at t = 0). At t = 11 ms, the electrophoretic conductivity can be determined from the Nerst-Planck equation [52]. For better comparison, only the increase in conductivity is shown, i.e., the conductivity is zero at t = 11 ms. The difference in conductivity with respect to the applied field is mainly due to the variation in the number of pores formed per cell and their average radius, with higher conductivity for higher pore number and radius. The difference in the ionic efflux with different E-fields of a single pair of inner/outer pores is less significant in the nanosecond regime. As can be seen in Figure 13, the conductivity is higher with the square wave pulse than the corresponding exponential decay pulse, due to the initial E(0) = 18 kV/cm field being sustained for the entire 11 ns period. Due to cytoplasmic leakage present after the pulse, *E coli* electroporation is not a fully reversible process. Thus, two successive pulses under crossed polarization are not possible [19].

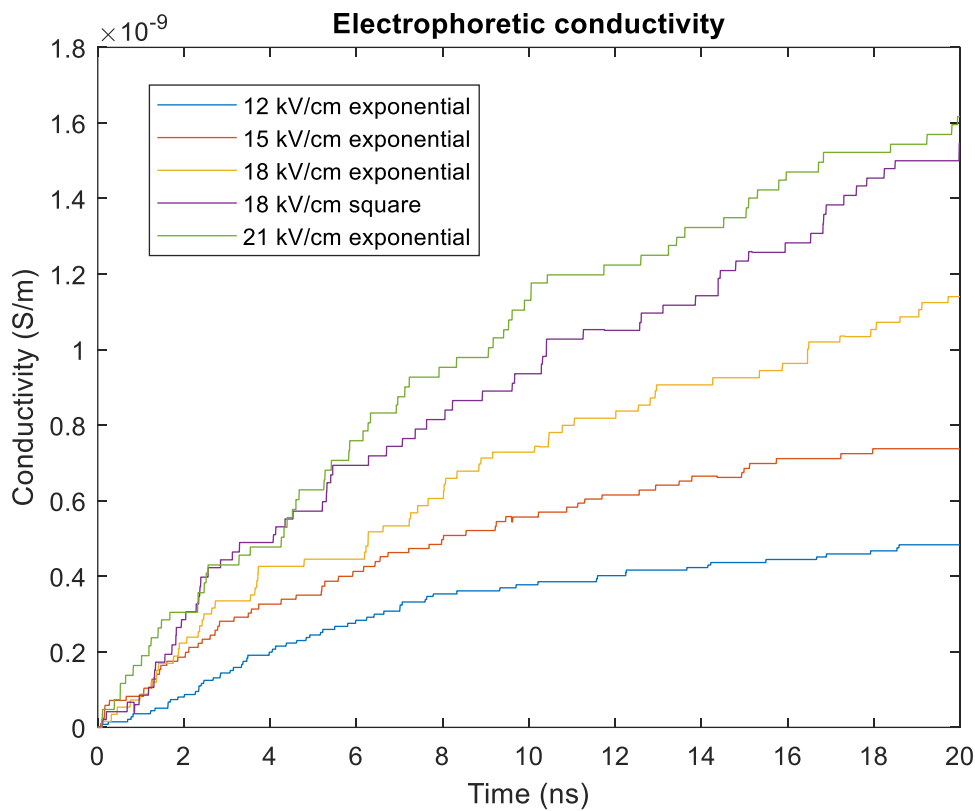

**Figure 13.** Simulation results probing the effect of changes in the pulse electric field amplitude on the temporal increase in the conductivity of electrocompetent *E. coli*. Time t = 0 corresponds to 11 ms after the pulse rising time. The starting conductivity of zero corresponds to the conductivity of the sample at 11 ms.

**DISCUSSION**

The Brownian dynamics approach has been chosen over the MD method (microscopic) and the continuum model (macroscopic) because it is computationally fast while at the same time the level of abstraction is low enough so that effects can be reasonably measured. From the BD model calculations, it was found that the main driving force of ionic uptake during the electric pulse is the one due to the externally applied electric field (electrophoretically mediated transfer), while other effects such as random collisions of ions with the medium or induced charges at the membrane only decrease ion throughput slightly. In Eqn. (1), the order of magnitude for the Coulomb and repulsive forces depends on the distance amongst the ions; the larger the distance the smaller the force. On average, the short-range repulsive force is several orders of magnitude smaller than the other forces ( < $10^{-17}$ N), while Coulomb, induced image, random and friction forces are in the order of $10^{-10}$ N. The applied electric field produces a force outside the membrane ~$2.8 \times 10^{-13}$ N. However, inside the membrane it becomes stronger and comparable to the Coulomb force. The main driving force is due to the applied electric field since it is oriented in one direction, while other forces point in all directions and cancel out or have their overall magnitude greatly diminished.

When applying the PEF the main contributors to the conductivity (Figure 11) are the $K^+$ cations due to their higher intra-cellular concentration, small size, and high diffusivity. $Mg^{2+}$ cations also contribute significantly to the conductivity due to their double charge and relatively high concentration. After the pulse, however, $Mg^{2+}$ contribute less due to their bigger size and thus lower diffusion. Despite having positive charges, few $Ca^{2+}$ and $Na^+$ ions are transported during the pulse because of low concentration and diffusivity (or mobility). The negatively charged $Cl^-$ anions move slowly in the opposite direction to the field, with strong attraction by the positive ions.

The bulk diffusion coefficient and conductivity calculations were compared to experimental observations, constituting a framework for calibrating the variables and validating the model in order to obtain realistic values for the unknown transport parameters, i.e., the mobility, diffusion, and translation time of ions within the pores. With the external electric field set to correspond to the applied pulse, the BD model calculates the conductivities due to electrophoretic transport of ions. With the electric field set to zero, the model calculates the increase in conductivity due to passive diffusion. These results present a tendency which is integrated into an analytical solution of the Nernst-Planck equation (Eqn. 40) for comparison with the experimentally determined extracellular conductivity. Measurements were taken over several seconds after the pulse when ion transport is due to facilitated and simple passive diffusion through long lived pores and protein channels [52]. The model fits reasonably well to the experiments considering its limitations (Figure 12). Further validations could be carried out by comparison of results with other microscopic and macroscopic models. The experimental patch-clamp technique can better validate the model with measurements taken in the microsecond time scale. Other experimental techniques, such as voltage sensitive dyes, Raman spectroscopy and confocal microscopy, as well as the giant unilamellar vesicle model system might also help validate the model, as in the work of Sozer et. al. for MD simulation validation [61]. The primary measurements for comparison are conductivity, diffusion, mobility, crossing time, electric field and voltage disturbances at different temperatures within the electroporated region.

A physical approximation introduced by the BD simulation is the neglect of the hydration (or solvation) spheres. Typically, the presence of polar water molecules leads to a shielded arrangement around the ions, which can best be captured by full MD simulations. This could become an issue for transport of ions across narrow pores, as the radius of the hydration shell approaches the channel opening. In order to incorporate in the BD model the average energetic contribution of the solute-solvent hydrogen bonds in the first solvation shell, implicit solvation models may be utilized [42,43]. Certain calculations, such as those involved in Green's function for the image charge forces, require summations to infinity. For practical purposes, a cut-off length has been introduced (.e., m,n = 4). Since the numbers used on a computer have a finite amount of digits, only a small subset of the rational numbers can be represented exactly with the IEEE binary floating-point arithmetic standard. Both rounding up and down introduce several errors in the calculations, such as the cancelation of significant digits due to subtraction of nearly equal number, division by a number with small magnitude, or multiplication by a large magnitude. Whenever possible, polynomials and equations have been written in forms that minimize the number of these calculations in order to make the occurrence of round-off errors less probable.

In order to better simulate pore formation and closure, more membrane details could be included in the model. Although the exact structure of pores is unknown experimentally, adding the effects of hydrophobic and hydrophilic molecules that form the membrane's lipid bilayer might help get more refined simulation results. To this aim, a hybrid MD-BD simulation scheme may be used which includes dipole shifts and configurational changes. Poisson solvers may be utilized to obtain the potential energy within the membrane and pore for a given configuration of ions. Changes in the resting membrane potential due to the ionic gradient may also be calculated using the Goldman-Hodkin Katz equation [7].

Potential future applications of the BD numerical simulation scheme include the study of cell death caused by ion dysregulation during electroporation. The model could be extended to simulate transport through proteins, e.g., voltage-gated ion channels as modeled in work reviewed by Sansom et. al. [57], as well as the transport of molecular chains. This will be an important step towards predicting the electric-field assisted entry and insertion of macromolecules such as polymers and DNA into cells.

## ACKNOWLEDGMENTS

This research work was funded by Consejo Nacional de Ciencia y Tecnología (CONACYT-Paraguay), grant number 14-INV-189. The authors would like to thank the Gustave Roussy Institute for allowing the use of equipment, cells and reactants which were used in conducting the experimental validation of the model.

## APPENDIX

*Derivations of induced image charges at the pores*

Green's function of the interior of a cylinder of radius $\rho_0$ and length $2z_0$ [64][65] is given by:

$$G(\rho, \emptyset, z, \rho', \emptyset', z')$$
$$= \frac{1}{2\pi\varepsilon_0\rho_0} \sum_{m=0}^{\infty} \sum_{n=1}^{\infty} \left\{ \frac{\epsilon_m \cos[m(\emptyset - \emptyset')] J_m(j_{m,n}\rho) \; J_m(j_{m,n}\rho')}{J_{m,n}J_{m+1}^2(j_{m,n})} \frac{\sinh[j_{m,n}(\alpha + z_<)]\sinh[j_{m,n}(\alpha - z_>)]}{\sinh(2j_{m,n}\alpha)} \right\}$$

(A1)

where $\epsilon_m = 2 - \delta_{m,0}$. $\delta_{0,0} = 1$ and $\delta_{m,0} = 0$ otherwise. $j_{m,n}$ is the $n^{th}$ zero of the Bessel function $J_m(x)$; $\alpha = z_0/\rho_0$ is the aspect ratio of the cylinder. The $\rho$ and $z$ cylindrical coordinates are scaled by $\rho_0$ and their origin is the center of the cylinder. The coordinate $z_<$ is the lesser of $z$ and $z'$, and $z_>$ is the greater.

The $z$ component on a charge $q$ can be found by evaluating Green's function $G$ (Eqn. A1) at $\rho' = \rho, \emptyset' = \emptyset, z' = z$ and taking the gradient following Smythe [60]:

$$F_z = -\frac{q^2 \nabla G(\rho,\emptyset,z,\rho',\emptyset',z')}{2} = \frac{q^2}{2\pi\varepsilon_0\rho_0} \sum_{m=0}^{\infty} \sum_{n=1}^{\infty} \frac{\epsilon_m J_m^2(j_{m,n}\rho')}{J_{m+1}^2(j_{m,n})} \frac{\sinh[2j_{m,n}z']}{\sinh[2j_{m,n}\alpha']}.$$

(A2)

Since this approach yields a divergent series when calculating the radial force $F\rho$, a different technique by Xiang *et. al.* (Xia93) is followed using an alternate form for the Green's function:

$$G(\rho, \emptyset, z, \rho', \emptyset', z') = \frac{1}{2\pi\varepsilon_0 z_0} \sum_{m=0}^{\infty} \sum_{n=1}^{\infty} \left\{ \epsilon_m \cos[m(\emptyset - \emptyset')] \sin\left[\frac{n\pi(z+1)}{2}\right] \sin\left[\frac{n\pi(z'+1)}{2}\right] T_n(\rho, \rho') \right\}$$

(A3)

where

$$T_n(\rho, \rho') = \frac{I_m\left(\frac{n\pi\rho_>}{2}\right)}{I_m\left(\frac{n\pi}{2\alpha}\right)} \left[ I_m\left(\frac{n\pi}{2\alpha}\right) K_m\left(\frac{n\pi\rho_<}{2}\right) - K_m\left(\frac{n\pi}{2\alpha}\right) I_m\left(\frac{n\pi\rho_<}{2}\right) \right],$$

(A4)

$I_m$ y $K_m$ are hyperbolic Bessel functions. Scaling is by $z_0$. The charge distribution $\sigma$ on the cylinder surface S can be found from Green{s function (Eqn. A3):

$$\sigma = -\varepsilon_0 q \nabla G(\rho, \emptyset, z, \rho', \emptyset', z'),$$

(A5)

where the gradient is taken with respect to $\rho, \emptyset, z$. Using the identity

$$I_m(u)K'_m(u) - K_m(u)I'_m(u) = -1/u, \tag{A6}$$

the surface charge on the cylinder is:

$$\sigma_c = \frac{-q}{2\pi\varepsilon_0 z_0}\sum_{m=0}^{\infty}\sum_{n=1}^{\infty}\left\{\epsilon_m \cos[m(\emptyset-\emptyset')]\sin\left[\frac{n\pi(z+1)}{2}\right]\sin\left[\frac{n\pi(z'+1)}{2}\right]\frac{I_m\left(\frac{n\pi\rho'}{2}\right)}{I_m\left(\frac{n\pi}{2\alpha}\right)}\right\}. \tag{A7}$$

The Surface charge creates a pressure directed outwards of the conducting membrane pore:

$$p = \frac{\sigma^2}{2\varepsilon_0}. \tag{A8}$$

Thus, the force on the ion is given by:

$$F_\rho = \int_S \frac{\sigma^2 da}{2\varepsilon_0}. \tag{A9}$$

The radial force on the ion from the above equation is:

$$F_\rho = -\frac{\rho_0}{4\pi\varepsilon_0}\int_0^{z_0} dz \int_0^{2\pi} d\emptyset\, 2\pi\sigma^2 \cos(\emptyset - \emptyset'), \tag{A10}$$

Considering the identity:

$$\int_0^{2\pi} \cos(x)\cos(mx)\cos(nx)\,dx = \begin{cases} \pi & \text{for } m=0 \text{ and } n=1 \\ \pi & \text{for } m=1 \text{ and } n=0 \\ \frac{\pi}{2} & \text{for } m-n=\pm 1 \text{ and } n,m>0 \\ 0 & \text{otherwise} \end{cases}, \tag{A11}$$

the radial force becomes [70]:

$$F_\rho = \frac{q^2}{2\pi\varepsilon_0\rho_0 z_0}\sum_{m=0}^{\infty}\sum_{n=1}^{\infty}\left\{\sin^2\left[\frac{n\pi(z+1)}{2}\right]\sin\left[\frac{n\pi(z'+1)}{2}\right]\frac{I_m\left(\frac{n\pi\rho'}{2}\right)I_{m+1}\left(\frac{n\pi\rho'}{2}\right)}{I_m\left(\frac{n\pi}{2\alpha}\right)I_{m+1}\left(\frac{n\pi}{2\alpha}\right)}\right\}. \tag{A12}$$

**CONFLICT OF INTEREST**

No benefits in any form have been or will be received from a commercial party related directly or indirectly to the subject of this manuscript.

# REFERENCES

1. Abrahamson, A. A. Born-Mayer-type interatomic potential for neutral ground-state atoms with Z=2 to Z=105. *Phys. Rev.* 178:76–79, 1969.
2. Alegun, O., A. Pandeya, J. Cui, I. Ojo, and Y. Wei. Donnan potential across the outer membrane of Gram-negative bacteria and Its effect on the permeability of antibiotics. *Antibiot. Basel* 10:701, 2021.
3. Basavaraju, M., and B. S. Gunashree. Escherichia coli: An Overview of Main Characteristics. In: Escherichia coli - Old and New Insights. IntechOpen, 2023.
4. Bayer, M. E. Zones of membrane adhesion in the cryofixed envelope of Escherichia coli. *J Struct Biol* 107:268–280, 1991.
5. Berendsen, H. J. Molecular Dynamics Simulations: The Limits and Beyond. In: Computational Molecular Dynamics: Challenges, Methods, Ideas. Lecture Notes in Computational Science and Engineering. , Berlin, Heidelberg: Springer, 1999.
6. Birdsall, C. K., and A. B. Langdon. Plasma Physics via Computer Simulations. McGraw-Hill Book Company, 1985, 56 pp.
7. Bonzanni, M., S. L. Payne, M. Adelfio, D. L. Kaplan, M. Levin, and M. J. Oudin. Defined extracellular ionic solution to study and manipulate the cellular resting membrane potential. *Biol Open* 9:, 2020.
8. Catipovic, M. A., P. M. Tyler, J. G. Trapani, and A. R. Carter. Improving the quantification of Brownian motion. *Am. J. Phys.* 81:485–491, 2013.
9. Chen, J. C., and A. S. Kim. Brownian Dynamics, Molecular Dynamics, and Monte Carlo modeling of colloidal systems. *Adv Colloid Interface Sci* 112:159–73, 2004.
10. Chini, T. K., and D. Ghose. On the interaction potential in low energy ion scattering. *Nucl. Instrum. Methods Phys. Res. Sect. B* 42:293–294, 1989.
11. Chung, S., T. W. Allen, M. Hoyles, and S. Kuyucak. Permeation of ions across the potassium channel: Brownian dynamics studies. *Phys. Rev. E* 77:2117–2155, 1999.
12. Chung, S., M. Hoyles, T. Allen, and S. Kuyucak. Study of ionic currents across a model membrane channel using Brownian dynamics. *Biophys. J.* 75:793–809, 1998.
13. Corry, B., T. W. Allen, S. Kuyukak, and S.-H. Chung. Mechanisms of Permeation and Selectivity in Calcium Channels. *Biophys. J.* 80:195–214, 2001.
14. Cory, B., M. Hoyles, T. W. Allen, M. Walker, S. Kuyucak, and S. Chung. Reservoir boundaries in Brownian dynamics simulations of ion channels. *Biophys. J.* 82:1975–1984, 2002.
15. DeBruin, K. A., and W. Krassowska. Modeling electroporation in a single cell. I. Effects Of field strength and rest potential. *Biophys J* 77:1213–24, 1999.
16. Diem, M., and C. Oostenbrink. The effect of different cutoff schemes in molecular simulations of proteins. *J. Comput. Chem.* 41:, 2020.
17. Einstein, A. On the movement of small particles suspended in a stationary liquid demanded by the molecular kinetic theory of heat (original in german). *Ann Phys* 17:549–560, 1905.
18. El-Hag, A., and S. H. Jayaram. Effect of biological cell size and shape on killing efficiency of pulsed electric field. , 2008.
19. Eynard, N., F. Rodriguez, J. Trotard, and J. Tessie. Electrooptics studies of Escherichia coli electropulsation: orientation, permeabilization, and gene transfer. *Biophys J* 75:2587–96, 1998.
20. Gapsys, V., and B. Groot. On the importance of statistics in molecular simulations for thermodynamics, kinetics and simulation box size. *eLife* 9:e57589, 2020.
21. Gothelf, A., L. M. Mir, and J. Gehl. Electrochemotherapy: results of cancer treatment using enhanced delivery of bleomycin by electroporation. *Cancer Treat Rev* 29:371–387, 2003.
22. Graham, L. L., T. J. Beveridge, and N. Nanninga. Periplasmic space and the concept of the periplasm. *Trends Biochem Sci* 16:328–329, 1991.

23. Green, M. S. Markoff Random Processes and the Statistical Mechanics of Time-Dependent Phenomena. II. Irreversible Processes in Fluids. *J Chem Phys* 22:398–413, 1954.
24. van Gunsteren, W. F., and H. J. Berendsen. Algorithms for Brownian dynamics. *Mol. Phys.* 45:637–647, 1982.
25. Hastings, C. Approximations for digital computers. Princeton, NJ: Princeton Univ. Press, 1955.
26. Ho, S. Y., and G. S. Mittal. Electroporation of cell membranes: A review. *Crit Rev Biotechnol* 16:349–362, 1996.
27. Hoejholt, K. L., T. Muzic, S. D. Jensen, L. T. Dalgaard, M. Bilgin, J. Nylandsted, T. Heimburg, S. K. Frandsen, and J. Gehl. Calcium electroporation and electrochemotherapy for cancer treatment: Importance of cell membrane composition investigated by lipidormics, calorimetry and in vitro efficacy. *Sci Rep* 9:4758, 2019.
28. Hoyles, M., S. Kuyucak, and S. Chung. Computer simulation of ion conductance in membrane channels. *Phys. Rev. E* 58:3654–3661, 1998.
29. Hu, Q., and R. P. Joshi. Comparative evaluation of transmembrane ion transport due to monopolar and bipolar nanosecond, high-intensity electroporation pulses based on full three-dimensional analyses. *J. Appl. Phys.* 122:034701, 2017.
30. Idalia, V.-M. N., and F. Bernardo. Escherichia coli as a Model Organism and Its Application in Biotechnology. In: Escherichia coli – Recent Advances on Physiology, Pathogenesis and Biotechnological Applications. InTechOpen, 2017.
31. Im, W. S., S. Seefeld, and B. Roux. A grand canonical Monte Carlo - Brownian dynamics algorithm for simulating ion channels. *Biophys. J.* 79:788–801, 2000.
32. Joshi, R. P., Q. Hu, R. Aly, K. H. Schoenbach, and H. P. Hjalmarson. Self-consistent simulations of electroporation dynamics in biological cells subjected to ultrashort electrical pulses. *Phys. Rev. E* 64:011913, 2001.
33. Kestin, J., M. Sokolov, and W. Wakeham. Viscosity of liquid water in the range −8 °C to 150 °C. *J. Phys. Chem. Ref. Data* 7:941, 1978.
34. Kotnik, T., L. Rems, M. Tarek, and D. Miklavcic. Membrane electroporation and electropermeabilization: mechanisms and models. *Annu. Rev. Biophys.* 6:63–91, 2019.
35. Krassowska, W., and P. D. Filev. Modeling electroporation in a single cell. *Biophys. J.* 92:404–417, 2007.
36. Kubo, R. Statistical-Mechanical Theory of Irreversible Processes. I. General Theory and Simple Applications to Magnetic and Conduction Problems. *J Phys Soc Jpn* 12:570–586, 1957.
37. Lemons, D. S., and A. Gythiel. Paul Langevin's 1908 paper "On the Theory of Brownian Motion" ["Sur la theorie du mouvement brownien," C. R. Acad. Sci. (Paris) 145, 530-533 (1908)]. *Am J Phys* 65:1079–1081, 1997.
38. Leontiadou, H., A. E. Mark, and S. J. Marrink. Ion transport across transmembrane pores. *Biophys. J.* 92:4209–4215, 2007.
39. Levitt, D. G. Modeling of ion channels. *J Gen Physiol* 113:789–94, 1999.
40. Lide, D. R., and H. V. Kehiaian. CRC Handbook of Thermophysical and Thermochemical Data. Boca Raton, FL: CRC Press, 1994.
41. Lo, C. J., M. C. Leake, T. Pilizota, and R. M. Berry. Nonequivalence of membrane voltage and ion-gradient as driving forces for the bacterial flagellar motor at low load. *Biophys J* 93:294–302, 2007.
42. Lomize, A. L., I. D. Pogozheva, and H. I. Mosberg. Anisotropic solvent model of the lipid bilayer. 1. Parameterization of long-range electrostatics and first solvation shell effects. *J. Chem. Inf. Model.* 51:918–29, 2011.
43. Lomize, A. L., I. D. Pogozheva, and H. I. Mosberg. Anisotropic solvent model of the lipid bilayer. 2. Energetics of insertion of small molecules, peptides, and proteins in membranes. *J. Chem. Inf. Model.* 51:930–46, 2011.

44. Miklavcic, D., and R. V. Davalos. Electrochemotherapy (ECT) and irreversible electroporation (IRE) - advanced techniques for treating deep-seated tumors based on electroporation. *Biomed Eng Online* 14:, 2015.
45. Milo, R., and R. Phillips. Cell Biology by the numbers. Garland Science, 2015, 400 pp.
46. Moran, J. L., N. N. Dingari, P. A. Garcia, and C. R. Buie. Numerical study of the effect of soft layer properties on bacterial electroporation. *Bioelectrochemistry* 123:261–272, 2018.
47. Neu, J., and W. Krassowska. Modeling postshock evolution of large electropores. *Phys. Rev. E* 67:021915, 2003.
48. Neu, J., K. Smith, and W. Krassowska. Electrical energy required to form large conducting pores. *Bioelectrochemistry* 60:107–114, 2003.
49. Papoulis, A. Probability, random variables, and stochastic processes. New York: McGraw Hill, 1984.
50. Piggot, T., D. Holdbrook, and S. Khalid. Electroporation of the E coli and S. Aureus membranes: Molecular Dynamics simulations of complex bacterial membranes. *J. Phys. Chem. B* 115:13381–13388, 2011.
51. Pluhackova, K., and R. A. Böckmann. Biomembranes in atomistic and coarse-grained simulations. *J Phys Condens Matter* 27:323103, 2015.
52. Pucihar, G., T. Kotnik, D. Miklavcic, and J. Tessie. Kinetics of transmembrane transport of small molecules into electropermeabilized cells. *Biophys. J.* 95:2837–2848, 2008.
53. Riley, M. Correlates of Smallest Sizes for Microorganisms. , 1999.
54. Rosazza, C., S. Haberl Meglic, A. Zumbusch, M. Rols, and D. Miklavcic. Gene electrotransfer: A mechanistic perspective. *Curr. Gene Ther.* 16:98–129, 2016.
55. Sabah, N. H. Rectification in biological membranes. *IEEE EMB Mag.* 106–113, 2000.
56. Sandblom, J., P. George, and J. Galvanovskis. The frequency response of ion channel currents in a combined diffusion and barrier type model. *J. Theor. Biol.* 176:153–160, 1995.
57. Sansom, M., I. Shrivastaya, J. Bright, J. Tate, C. Capener, and P. Biggin. Potassium channels: structures, models, simulations. *Biochim. Biophys. Acta BBA - Biomembr.* 1565:11, 2002.
58. Saulis, G., and R. Saule. Size of the pores created by an electric pulse: microsecond vs millisecond pulses. *Biochim Biophys Acta* 1818:3032–9, 2012.
59. Shen, L. C., and J. A. Kong. Applied Electromagnetism. Boston, MA: PWS Publishers, 1987.
60. Smythe, W. R. Static and Dynamic Electricity. New York: McGraw-Hill, 1968.
61. Sözer, E. B., S. Haldar, P. S. Blank, F. Castellani, P. T. Vernier, and J. Zimmerberg. Dye Transport through Bilayers Agrees with Lipid Electropore Molecular Dynamics. *Biophys. J.* 119:1724–1734, 2020.
62. Spanggaard, I., M. Snoj, A. Cavalcanti, C. Bouquet, G. Sersa, C. Robert, M. Cemazar, E. Dam, B. Vasseur, P. Attali, L. M. Mir, and J. Gehl. Gene electrotransfer of plasmid antiangiogenic metargidin peptide (AMEP) in disseminated melanoma: safety and efficacy results of a phase I first-in-man study. *Hum Gene Ther Clin Dev* 24:99–107, 2013.
63. Swope, W. C., H. C. Andersen, P. H. Berens, and K. R. Wilson. A computer simulation method for the calculation of equilibrium constants for the formation of physical clusters of molecules: Application to small water clusters. *J. Chem. Phys.* 76:648, 1982.
64. Tinkle, M. D., and S. E. Barlow. Workshop on Non-Neutral Plasmas. , 2001.
65. Tinkle, M. D., and S. E. Barlow. Image charge forces inside conducting boundaries. *J. Appl. Phys.* 90:1612–1624, 2001.
66. Turq, P., F. Lantelme, and H. L. Firedman. Brownian dynamics: its application to ionic solutions. *J. Chem. Phys.* 66:3039–3044, 1977.
67. Verlet, L. Computer "Experiments" on Classical Fluids. I. Thermodynamical Properties of Lennard–Jones Molecules. *Phys. Rev.* 159:98–103, 1967.
68. Weaver, J. C. Electroporation of cells and tissues. *IEEE Trans Plasma Sci* 28:24–33, 2000.

69. Weaver, J. C., K. C. Smith, A. T. Esser, R. S. Son, and T. R. Gowrishankar. A brief overview of electroporation pulse strength-duration space: a region where additional intracellular effects are expected. *Bioelectrochemistry* 87:236–43, 2012.
70. Xiang, X., P. B. Grosshans, and A. G. Marshall. Image charge'induced ion cyclotron orbital frequency shift for orthorhombic and cylindrical FT'ICR ion traps. *Int. J. Mass Spectrom. Ion Process.* 125:33–43, 1993.
71. Young, H., and R. Freedman. Sears and Semansky's University Physics. New York, NY: Addison Wesley, 1996.